\documentclass[journal]{IEEEtran}
\usepackage{times}
\usepackage{color,soul}
 
\usepackage{multirow}
     
\ifCLASSINFOpdf
   \usepackage[pdftex]{graphicx}
\else
\usepackage[dvips]{graphicx}

 \usepackage{amssymb,amsmath}
\fi
  
\usepackage{gensymb}
\usepackage{enumerate}
\usepackage{subfig}
\usepackage{cite}
\usepackage{caption}  
\usepackage{url}
\usepackage{breakurl}
\usepackage[breaklinks]{hyperref}
\usepackage{array}
  
\begin{document}
 
\title{Smartphone-based Wellness Assessment Using Mobile Environmental Sensors}

\author{Katherine McLeod, \IEEEmembership{Student Member,~IEEE}, Petros Spachos,~\IEEEmembership{Senior Member, ~IEEE}, and Konstantinos~N.~Plataniotis,~\IEEEmembership{Fellow, ~IEEE}

\thanks{K. McLeod and P. Spachos are with the School of Engineering, University of Guelph, Guelph, ON, Canada (e-mail: kmcleo04@uoguelph.ca, petros@uoguelph.ca)}
\thanks{K. N. Plataniotis is with the Department of Electrical and Computer Engineering, University of Toronto, Toronto, ON, Canada (e-mail: kostas@ece.utoronto.ca)}
}

\bstctlcite{IEEEexample:BSTcontrol}

\maketitle

\begin{abstract}

Mental health and general wellness are becoming a growing concern in our society. Environmental factors contribute to mental illness and have the power to affect a person's wellness. This work presents a smartphone-based wellness assessment system and examines if there is any correlation with one's environment and their wellness.  The introduced system was initiated in response to a growing need for individualized and independent mental health care and evaluated through experimentation.  The participants were given an Android smartphone and a mobile sensor board and they were asked to complete a brief psychological survey three times per day.  During the survey completion,  the board in their possession is reading environmental data.  The five environmental variables collected are temperature, humidity, air pressure, luminosity, and noise level. Upon submission of the survey, the results of the survey and the environmental data are sent to a server for further processing. Three experiments with 62  participants in total have been completed.  The correlation most regularly deemed statistically significant was that of light and audio and stress.

\end{abstract}


\IEEEpeerreviewmaketitle

\section{Introduction}
\label{introduction}
\IEEEPARstart{M}{ental} wellness is a growing concern worldwide.  There is a need for solutions that can reach a diverse group of people fast and easy.  In the United States, a depressed mood is the leading cause of disability about people aged 15-44~\cite{urgent}. However, any potential solution should always consider the associated cost. In Canada, the costs related to mental health care were estimated at \$42.3 billion~\cite{lop}.   At the same time, the cost of the lack of mental health care can be expensive.  Disability-related to a depressed mood in the United States is estimated to account for a \$31 billion loss annually in terms of productivity~\cite{urgent}.   
 
Over time, technology has been used more and more often to solve issues in healthcare~\cite{islam, zhang}. Recent advantages in smartphones and sensor  technology can be a promising solution for a plethora of healthcare system challenges. The high computational power of smartphones along with  the low cost and easy to use of sensor boards can alleviate many of the challenges in the existing health care systems~\cite{yzheng, catarinucci, mcleod4}. At the same time, wireless technologies such as WiFi Halow and Bluetooth Low Energy~(BLE) can provide efficient communication between smartphones and mobile sensor boards.

In this work, we introduce a framework that can be used for wellness assessment. The framework consists of two main components: a mobile application and a mobile device, the SensorTag~\cite{ti3}.  The SensorTag is a mobile sensor board that communicates with a smartphone  via BLE. It contains ten sensors, four of which are used in this system to collect five environmental variables: temperature, humidity, air pressure, light, and noise.  The mobile application connects to the SensorTag via BLE and asks the participant to complete a brief psychological survey three times per day.  The survey includes questions from three existing psychological surveys, the timeframes of each having been modified: the Pittsburgh Sleep Quality Index (PSQI) measuring sleep quality, the Perceived Stress Scale (PSS) measuring stress, and Kessler's Psychological Distress Scale (K10) measuring distress.  During the time that the participant is responding to the survey, the SensorTag is reading data for the five environmental variables.  Upon submission of the survey by the participant, the results along with the sensor data are all sent to the server via a secure WiFi network.

A feasibility testing of the system, with 6 participants was presented in~\cite{globalsip}.  In this work, along with the introduced framework, the results of three experiments are presented. The experiments were conducted in three different seasons during the year to collect data under different environmental conditions and examine the stress levels of the participants. In total, in all the three experiments 62 participants completed the survey three times per day for 530 valid submissions.

The main contributions of this work are listed below:
\begin{itemize}
\item An experimental framework for wellness assessment is introduced. The framework combines a smartphone along and a mobile sensor board for data collection. 
\item An Android application for data collection was developed, while BLE was found to be the most energy-efficient technology for this study.
\item Experimental results of 62 participants are presented, from three experiments. The experiment took place during three different seasons, Fall, Winter, and Spring to examine the effect of the environmental conditions on the participants.
\item Correlation analysis between the collected data and three psychological questionnaires is performed. According to the results, there is a statistically significant correlation between light intensity and audio level and stress.
\end{itemize}

The rest of this paper is organized as follows: Section~\ref{related} provides a review of the related work.  Section~\ref{proposed} introduces the proposed system, and a description of the psychological surveys is in Section~\ref{surveys}. Section~\ref{method} discusses the experimental method for each experimental phase. Section~\ref{results} presents and discusses the results of each phase.  Section~\ref{con} concludes the work.


\section{Related Works} \label{related}

Technological advantages can be used to alleviate several challenges in  healthcare. In~\cite{park}, a study is presented for a system designed to assess upper limb tremor in patients with Parkinson's disease.  The system makes use of the gyroscope and accelerometer in the smartphone to develop a dataset that can accurately quantify a patient's hand tremors.  The system boasts a low cost due to the use of a smartphone as hardware.  It is also noninvasive and allows for independent use by the patient.  The dataset compares 25 Parkinson's patients to 20 healthy volunteers.  The study concluded with an 82\% classification accuracy for the patients and a 90\% classification accuracy for the volunteers.  

Among the tools being used often in healthcare  are smartphones~\cite{osmani, Alshurafa, Barnett} along with smartphone accessories~\cite{mahmud}  and Internet of Things (IoT) devices~\cite{baker, islam}.  A benefit achievable through many IoT-based solutions is the production of low-cost and low-power models~\cite{mc2}.  Many fields already have commercially viable IoT options, including smart parking, precision agriculture, and water usage management.  Health care, as an application of IoT, is a promising but still developing application.

In~\cite{bipolar}, a smartphone-based system is introduced to recognize depressive and manic states.  They explain that psychiatric care remains one of the only facets of health care to be measured objectively using physiological parameters and symptoms.  The system introduced uses a dataset of ten patients over twelve weeks to extract ``features corresponding to all disease-relevant aspects in behavior".  When all modalities were fused, the system achieved a 76\% recognition accuracy.

Mobile sensors are being used with increasing frequency to accurately evaluate a subject's well-being. In~\cite{multi}, a smartphone-based system was created to remotely monitor the symptoms, behavior, and physiology of psychiatric patients.  The purpose of this design was to create a low-budget approach to mental health care, a division of health care with consistently low funds.  Over 100 participants completed several psychiatric questionnaires to be repeated on a weekly basis.   An application was developed to make use of the sensors on a Samsung Galaxy S III.  The application records actigraphy levels, ambient light levels, social network activity, and participant physiology (blood pressure and temperature).  The initial results  clearly demonstrate consistency in the actigraphy and social networking levels of a healthy control compared to a participant with a diagnosed borderline personality disorder.  In~\cite{madison}, a framework is described which determines a correlation between wellness and environmental factors. In this case, the environmental factors studied are temperature, humidity, and luminosity, with a correlation determined between luminosity and general well-being over an experimental period with 23 participants. 

In~\cite{mit},  wearable sensors and smartphone usage are used to evaluate correlation to students' academic performance, self-reported sleep quality, self-reported stress, and self-reported mental health.  The study gave wearable sensors to 66 student participants over 30 days, achieving 1,980 days of data.  Several surveys were completed by participants before beginning, including the Pittsburgh Sleep Quality Index (PSQI), the Perceived Stress Scale (PSS), and the Mental Health Composite Scale (MCS).  Each participant wore a collection of wearable sensors on their wrist for the duration of the experiment: an accelerometer, light sensor, actigraphy monitor, skin temperature sensor, and skin conductance sensor.  Many associations were determined with classification accuracies ranging from 67-92\%.  PSQI groups were in turn related to sleep regularity, confirming a hypothesized relationship between high stress levels and low sleep quality.  In particular, the wearable sensors achieved close to a 90\% classification accuracy for PSQI, PSS, and MCS.

IoT has been identified by many as a potential solution for the increasing demands on health care systems since it has previously allowed for automation in many industries.  In~\cite{baker}, an end to end IoT health care system is proposed identifying the key components necessary for a functional model and introduces an array of technologies that fits the requirements.  

The disadvantage with the greatest risk is a security concern, and it is therefore very important to identify any possible security concerns in these systems.  In~\cite{islam}, they identify several security requirements specific to IoT health care systems.  Confidentiality ensures that no transferred medical data is accessible to intruders of the system.  The challenges in building an IoT security system with these requirements are also some of the benefits to the system at large: a low-power CPU device with low memory and low energy have inherent limitations in terms of design.  

In comparison with the related work, this work demonstrates the possibility to successfully compare environmental data collected via mobile sensors to wellness surveys.  The evident advantage of our experimental system is the sole focus on the correlation of mental wellness to five environmental variables, without concern for any extra data such as physical wellness.  Also, we examine ambient noise, an environmental factor not explored before.  Our experimental system uses a digital microphone to evaluate noise levels around the participant throughout the day and search for any correlation to the individual's overall wellness.  This was added to our work following the initial testing phase, replacing the movement variable.  Compared to our previous work described in~\cite{madison}, our system has the advantage of two additional variables: audio amplitude and barometric pressure.


\section{System Overview} \label{proposed} 
 
\begin{figure}[t!]
\centering
\includegraphics[width=\linewidth]{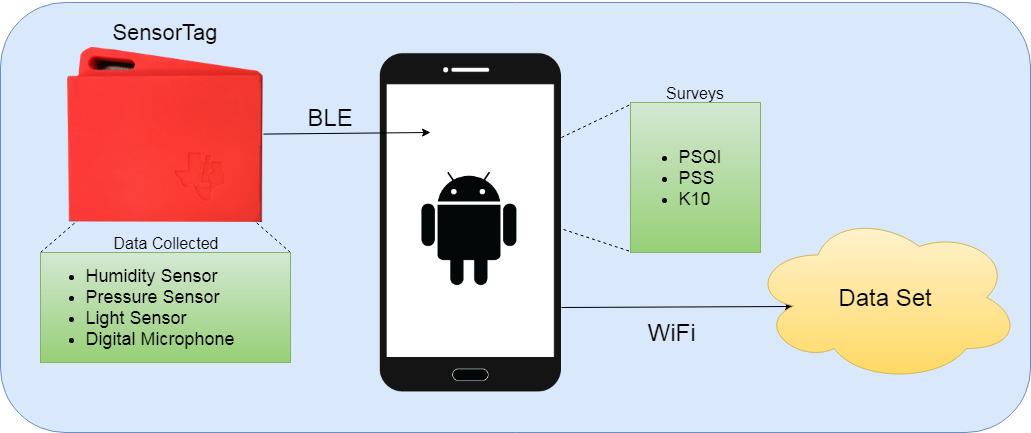}
\caption{Data collection from SensorTag to the Android application for a single user.}
\label{flow} 
\end{figure}

The framework has three main components: i) the smartphone with the mobile application, which communicates with both ii) the SensorTag BLE device and iii) the cloud server.  A general flowchart of the system can be seen in Fig.~\ref{flow}. In this
diagram, the SensorTag's four sensors that are used in this system are highlighted. The humidity/ temperature, pressure, light sensors, and
the microphone collect raw data for our five environmental variables: ambient temperature, humidity, air pressure, luminosity,
and ambient audio amplitude. The SensorTag communicates the raw data to the Android application via BLE. The application
is designed to provide questions from three pre-selected psychological surveys. The survey results in conjunction with
the raw sensor data are sent to the server over WiFi, for processing and storage. 

\subsection{Hardware requirements}

Following is a brief description of the hardware components and their specifications that were given to each participant in the study for use during the experimental period.

\subsubsection{Smartphone}
The smartphone used in the experiment should have BLE functionality and WiFi connectivity. For our experiments, the LG Nexus 5 smartphone was used, running Android 6.0.1. Three processes take place at the smartphone. First, it notifies the user when they should answer the surveys. Then, it collects the environmental data from the sensor board over BLE. Finally, it uploads all the data to the server when WiFi connectivity is available.

\subsubsection{Sensor board}

\begin{figure}[t!]
\centering
\includegraphics[width=0.9\linewidth]{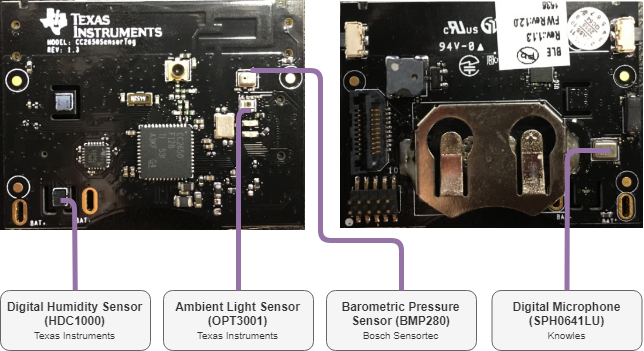}
\caption{Texas Instruments SensorTag with relevant sensors.}
\label{fig:hw1} 
\end{figure}

The SimpleLink BLE SensorTag was used for this experiment.  It is a small, portable collection of sensors~\cite{ti3}.  The model of SensorTag used for this experiment contains a CC2650 wireless MCU, which provides significantly low power consumption from the 3V coin cell battery and is compatible with Android.  SensorTag has ten sensors on board, while the five listed are used in the experiment and the rest are deactivated to save energy.  The sensors are shown in Fig.~\ref{fig:hw1}. In each experimental phase some of the five sensors are used, as indicated below:

\begin{itemize}
  \item Humidity/ Temperature Sensor: For our framework, the humidity/ temperature sensor measures two environmental variables: humidity and ambient temperature.  
  
  \item Barometric Pressure Sensor: The barometric pressure sensor is an absolute barometric pressure sensor, designed for use with mobile applications and low power consumption.
  
  \item Ambient Light Sensor:   This sensor is specifically designed for systems that involve human interactions with light, making it  ideal for measuring the ambient light of the participants' surroundings in the proposed system.  It is important that the participants ensure that the light sensor is positioned such that it is exposed to ambient during the data collection.

  \item Digital Microphone: The SPK0833 digital microphone~\cite{ti3} is the ambient audio sensor component of the SensorTag. The microphone is used to sense the magnitude of ambient sound in the participants' surroundings, regardless of the source of the noise.  This sensor was not used in the initial testing phase but was used in the experimental period.
    \end{itemize}
    
The 9-axis motion tracking was used in the initial testing phase~\cite{globalsip}, to track the participant's physical activity throughout the day in terms of magnitude and duration.  However,  it was not included in the three experiments. After the initial testing, we conclude that having the device in motion would interfere with the ability to directly read data to other sensors, particularly the ambient light sensor.

\subsubsection{Server}
The server is used for data collection and to perform the necessary correlations. It is important to make sure of a secure connection between the server and the smartphone. In our experiments, the smartphones can access the server only over a secure network.

\subsection{Mobile application}

\begin{figure}[t!]
\centering
\includegraphics[scale=0.4]{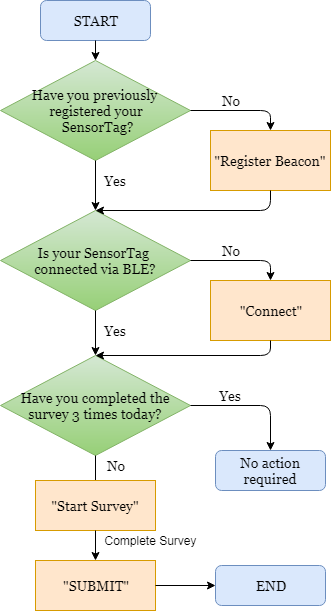}
\caption{Flow chart detailing one session with the mobile application.}
\label{fig:appflow} 
\end{figure}

A mobile application was developed during the initial experimental period~\cite{globalsip}.  The application was modified following the initial testing phase, but the main functionality remained the same.  The process of using the application for one survey session is described in the flow chart in Fig.~\ref{fig:appflow}.

The survey need only be completed three times per day, at which point no further action is required until the next day. Once the participant is ready to begin a survey, they select the ``Start Survey" button on the home screen.  As long as they are already registered and connected to the SensorTag, this will take them to the survey screen.  Then, the user has to answer each of the 33 survey questions provided.  Once each question has been answered, the user can submit the answers.  If any questions remain unanswered, the user will be notified to complete the entire survey.  When all questions contain an answer, the survey results and sensor data are sent to the server via WiFi and the user is notified of a successful transaction before exiting the application.  In the case of an incomplete transaction due to connectivity issues, the user is given the opportunity to connect to WiFi before trying to submit the same results again.

 \subsection{Communication technologies}
 
 Several  wireless technologies and protocols exist to transfer the data between devices.  Each standard has its benefits and is, therefore, best suited to systems with certain characteristics.  SensorTag has the following wireless technologies available:
 
\begin{itemize}

    \item \textbf{BLE}:  BLE was designed for the purpose of lower power consumption than Classic Bluetooth.  This is ideal for experiments that will take place over an extended period of time.  Low power consumption is an advantage, however, the range is shorter than WiFi.

    \item \textbf{Sub-1GHz}: Sub-1GHz is another popular technology among IoT devices. Although the data rate is low, it can reach miles with minimum energy requirements.
    
        \item \textbf{WiFi (2.4GHz)}: WiFi is most commonly used in Wireless Local Area Networks (WLAN).  WiFi's high availability makes it ideal for use with the IoT and its devices. Its long range when indoors can be an advantage but also create interference.
   
  \end{itemize}
  
 All three communication technologies are strong options for this system given the following requirements: low cost, low power, small amounts of data. Considering the low transmission range that can minimize interference, along with the low energy requirements and the high availability in many smartphone devices, BLE was selected for our experiments. At the same time  WiFi, while it has a high power consumption, is used to send data from the smartphones to the server due to range, simplicity, and availability.

\section{Description of Psychological Surveys}\label{surveys}

The surveys were selected in order to inquire about a variety of aspects of an individual's current wellness, including stress and sleep.  The questions and possible responses have been modified from their original form to reflect the fact that we are only interested in the participant's wellness at the time they are completing our surveys, rather than an extended period of time.  The three surveys described were used for the initial testing phase and the experimental period.  Following the initial testing phase, an additional question was added: ``How many people are around you right now? (ie. in the same room)".  This question essentially provides a sixth environmental variable, without the use of a sensor: the presence of other humans in the participant's space.

The questions from the three official surveys have been modified for this experiment.  The PSQI, PSS, and Kessler Psychological Distress Scale (K10) present questions pertaining to an individual's wellness over the past 30 days.  However, this time frame is not relevant to our experiment.  For instance, all PSS questions begin ``In the last month,..." and the K10 questions begin with ``During the last 30 days,...".  For this reason, the questions have been modified to reflect the individual's immediate well-being.  These modifications are reflected in Tables~\ref{psqi}, \ref{k10}, and \ref{pss}.

\subsubsection{Pittsburgh Sleep Quality Index (PSQI)}

The PSQI~\cite{psqi} was selected and modified to achieve quantification of the participant's sleep quality the previous night.  Since the surveys are taken by the participant 2-3 times per day, this survey is only presented during the first submission of the day.  This is because the participant's answers will not change until the next morning.  Our modified PSQI consists of 17 questions inquiring into quality and duration of sleep, as shown in Table~\ref{psqi}. 

\begin{table}[t!] 
\centering
\caption{PSQI questions answered by participant.}
\label{psqi}
\begin{tabular}{ |p{5.2cm}|p{2.8cm}|  }
  \hline
 \textbf{Question}  & \textbf{Possible Responses}   \\
 \hline\hline
  When did you go to bed last night?&   8-8:30pm, 8:30-9pm etc.    \\
 \hline
  How many hours did you spend in bed last night?&   0-1 hour, 1-2 hours etc.    \\
 \hline
   How many hours of sleep did you get last night?&   0-1 hour, 1-2 hours etc.    \\
 \hline
 Last night, did you go to sleep within 30 minutes?   &   Yes/No \\
 \hline
 Last night, did you wake up in the middle of the night?   &   Yes/No \\
 \hline
 Last night, did you get up to use the bathroom?   &   Yes/No \\
 \hline
 Last night, did you have trouble breathing properly?   &   Yes/No \\
 \hline
  Last night, did you cough or snore loudly?   &   Yes/No \\
 \hline
 Last night, did you feel too cold?   &   Yes/No \\
 \hline
 Last night, did you feel too hot?   &   Yes/No \\
 \hline
 Last night, did you have bad dreams?   &   Yes/No \\
 \hline
 Last night, did you have pain?   &   Yes/No \\
 \hline
 Did you take medicine to help you sleep last night?   &   Yes/No \\
 \hline
 Have you had trouble staying awake in the past 24 hours?   &   Yes/No \\
 \hline
  Have you had trouble keeping enthusiasm to get things done in the past 24 hours?   &   Yes/No \\
 \hline
 How would you rate last night's sleep overall?   &   1, 2, 3, 4, 5 \\
 \hline
\end{tabular}%

\end{table}

\subsubsection{Perceived Stress Scale (PSS)}

The PSS~\cite{pss} was selected and modified to quantify the degree to which the participant is feeling stressed at the time of the survey.  Our modified PSS consists of 10 questions inquiring into the presence or lack of various symptoms of stress, which can be seen in Table~\ref{pss}.

\begin{table}[t!] 
\centering
\caption{PSS questions answered by participant.}
\label{pss}
\begin{tabular}{ |p{5.2cm}|p{2.8cm}|  }
 \hline

 \textbf{Question}  & \textbf{Possible Responses}   \\
 \hline\hline
  Do you feel upset by something that happened unexpectedly?&   Yes/No    \\
 \hline
  Do you feel unable to control the important things in your life?&   Yes/No    \\
 \hline
  Do you feel stressed?&   Yes/No    \\
 \hline
  Do you feel confident about your ability to handle your personal problems?&   Yes/No    \\
 \hline
  Do you feel that things are going your way?&   Yes/No    \\
 \hline
  Do you feel that you are able to cope with all the things you have to do?&   Yes/No    \\
 \hline
  Do you feel that you are able to control the irritations in your life?&   Yes/No    \\
 \hline
 Do you feel that you are on top of things?&   Yes/No    \\
 \hline
 Do you feel anger because of things that are outside of your control?&  Yes/No \\
 \hline
 Do you feel difficulties are piling up so high that you could not overcome them?   &   Yes/No \\
 \hline
\end{tabular}%

\end{table}

\subsubsection{Kessler Psychological Distress Scale (K10)}

The K10~\cite{k10} was selected and modified to quantify the participant's level of psychological distress at the time the survey is taken.  Our modified PSS consists of 10 questions pertaining to the relevance of various feelings of distress. For example, the survey asks whether the participant feels hopeless, nervous, or depressed.  The questions can be seen in Table~\ref{k10}.

\begin{table}[t!] 
\centering
\caption{K10 questions answered by participant.}
\label{k10}
\begin{tabular}{ |p{5.2cm}|p{2.8cm}|  }

  \hline
\textbf{Question}  & \textbf{Possible Responses}   \\
 \hline\hline
  Do you feel tired for no good reason?&   Yes/No    \\
 \hline
 Do you feel nervous?&   Yes/No    \\
 \hline
 Do you feel so nervous that nothing can calm you down?&   Yes/No    \\
 \hline
 Do you feel hopeless?&  Yes/No \\
 \hline
 Do you feel restless or fidgety?   &   Yes/No \\
 \hline
 Do you feel so restless that you can not sit still?&   Yes/No    \\
 \hline
 Do you feel depressed?&   Yes/No    \\
 \hline
 Do you feel that everything is an effort?&   Yes/No    \\
 \hline
 Do you feel so sad that nothing can cheer you up?&   Yes/No    \\
 \hline
 Do you feel worthless?&   Yes/No    \\
 \hline
\end{tabular}%

\end{table}

\section{Experimental methodology and correlation analysis} \label{method}
To evaluate the introduced framework, three experiments took place. The details of the experiments are shown in Table~\ref{exptable}. The first experiment with 8 participants took place in early  September, the second experiment with 20 participants took place in October through December  and the third in March. Conducting separate experiments with time between each was chosen, to analyze data from different seasons and to reflect different environments. Also, since the majority of the participant are undergraduate students  and stress and wellness is monitored in this experiment, to have a variation in the daily activities, the three experiments took place different time during the semester. The first experiment took place at the beginning of the semester, the second towards the end and the exam period, and the third in the middle of the term.

\begin{table}[t!] 
\centering
\caption{Information about the experiments.}
\label{exptable}
\begin{tabular}{ |l|c|c|c|}
\hline
 &  Participants &Dates  & Valid \\
  &  &  & submissions\\
  \hline
 \textbf{Experiment 1}  & 8 & Sept. 5 - Sept. 10  & 61\\
 \hline
 \textbf{Experiment 2} - & \multirow{2}{*}{10} & Oct. 15 - Oct. 20 & 50\\ \cline{3-4}
  \textbf{Group~A }      &  & Nov. 25 - Nov. 30   & 51\\
 \hline
 \textbf{Experiment 2 -} & \multirow{2}{*}{10}&  Nov.1 - Nov. 5  & 39\\\cline{3-4}
  \textbf{Group~B} &  & Dec.10 - Dec. 15  & 42\\

 \hline
 \textbf{Experiment 3}  &34  & March 1 -  March 20 & 287 \\
 \hline
\end{tabular}%
\end{table}
\subsection{Experimental procedure}
Prior to beginning the experiment, each participant was asked to attend a training session.  The objectives of the training were as follows:
    
    \begin{itemize}
  \item Educate the participant on their rights and the efforts to protect their privacy and data for the experiment, and have them sign a consent form denoting these points as approved by the University Research Ethics Board.
  
  \item Explain the expectations of them during the experiment and demonstrate the use of the application.
   
  \item Lend the participant a phone and SensorTag for the duration of the experiment.
  
  \end{itemize}
  
  When giving the participant instructions for their participation, the instructions were as follows:
  
  \begin{enumerate}
      \item Place the SensorTag on a flat surface such that the light sensor is oriented face up when ready to complete the survey.
      \item Turn on the SensorTag using the button on the side before opening the application.  Check it operates by making sure the green light is blinking.
      \item Check the sensor readings on the home screen of the application for reasonability.  The first sign that the SensorTag's battery is running low is zero readings from a sensor.  Contact the researcher if a replacement battery is needed.
      \item Complete the survey three times per day, when/ if on  campus.  Make sure multiple submissions in a day are several hours apart.  Must be connected to University WiFi to submit the survey.
  \end{enumerate}
  
  \begin{figure*}[t!] 
  \centering
  \captionsetup[subfloat]{farskip=0pt}%
\subfloat[Temperature]{\includegraphics[width=0.193\textwidth]{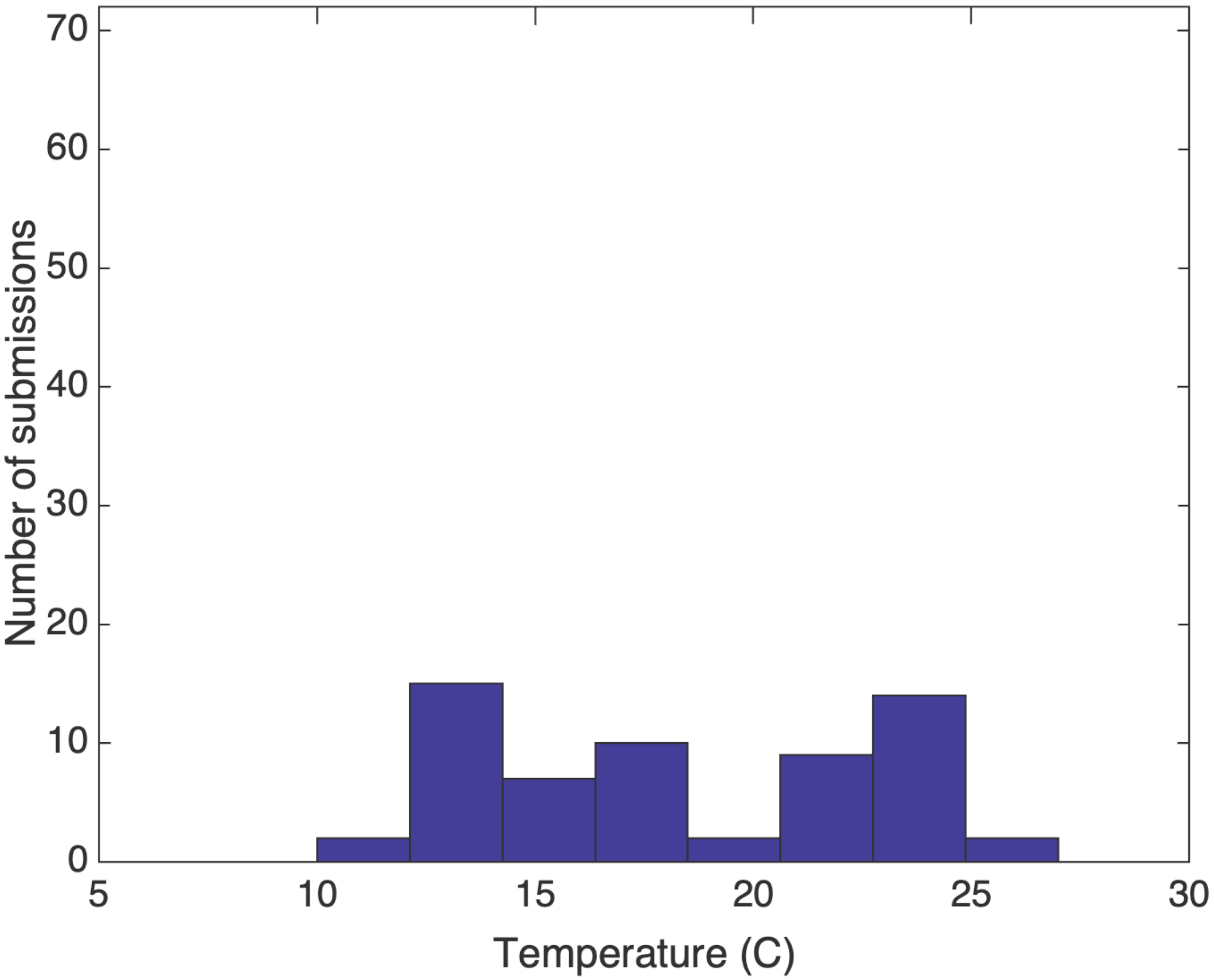}
    \label{fig:d1}}\hfill
\subfloat[Humidity]{\includegraphics[width=0.19\textwidth]{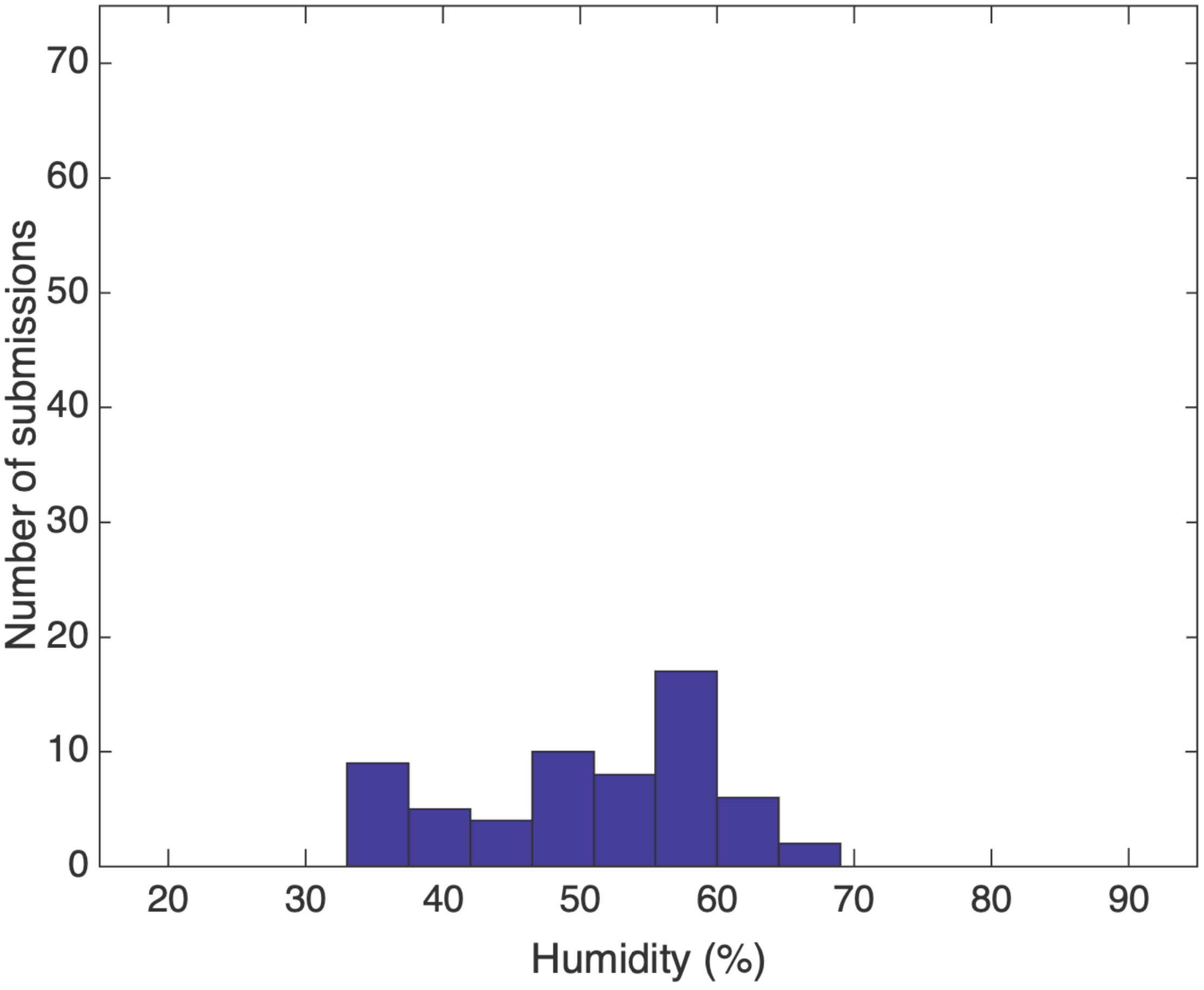}
    \label{fig:d2}}\hfill
\subfloat[Light]{\includegraphics[width=0.19\textwidth]{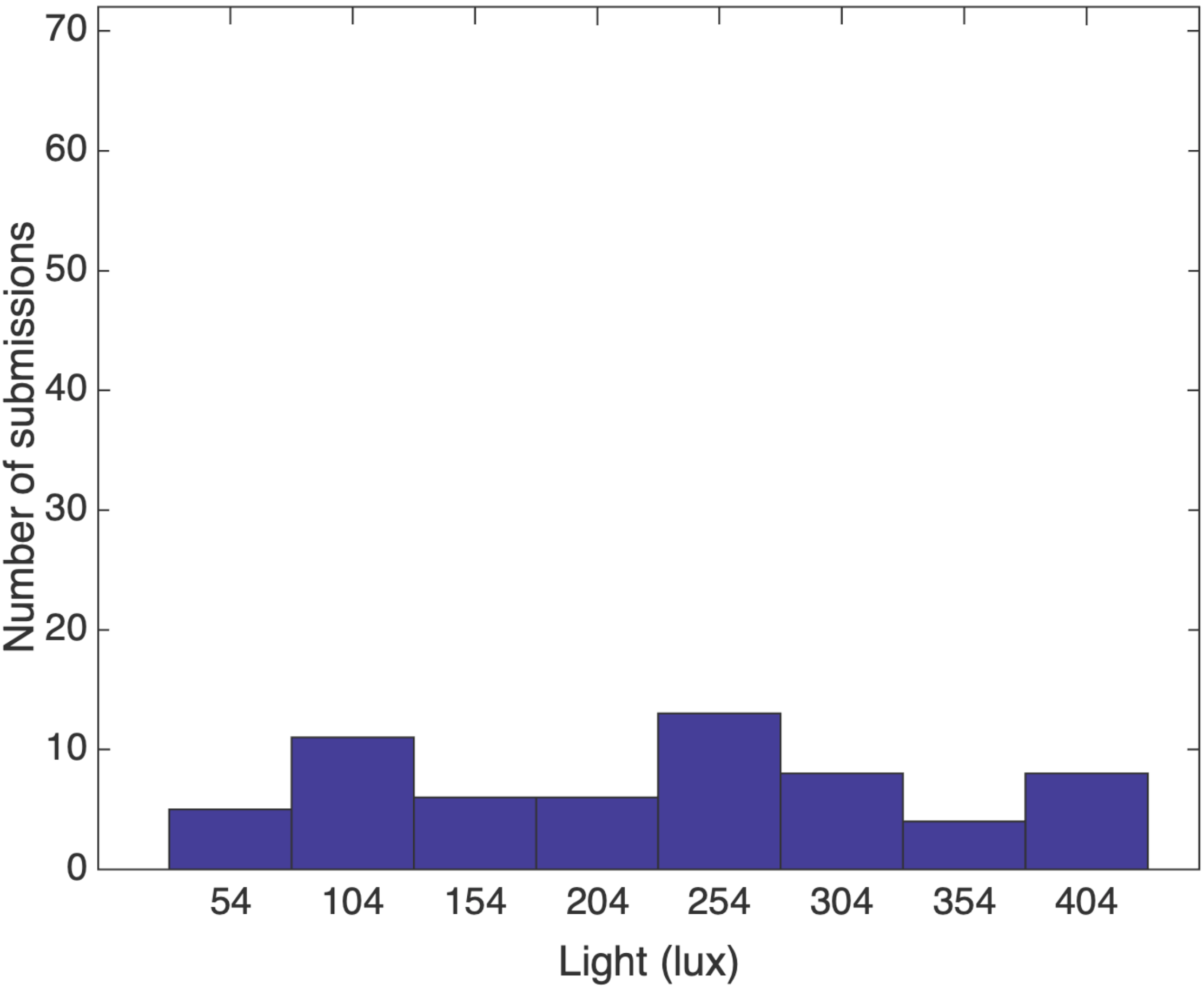}
       \label{fig:d3}}\hfill
 \subfloat[Pressure]{\includegraphics[width=0.193\textwidth]{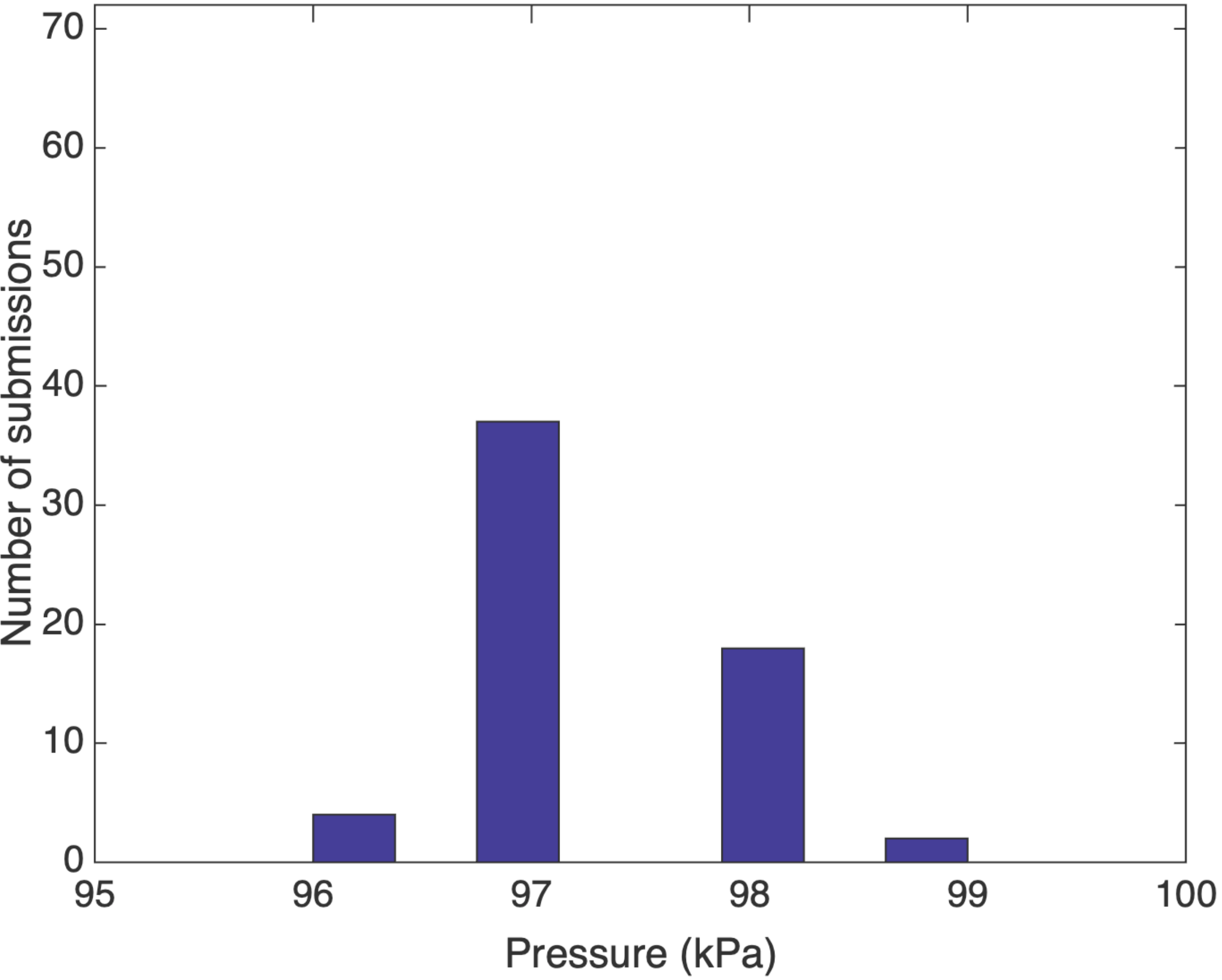}
    \label{fig:d4}}\hfill
\subfloat[Audio]{\includegraphics[width=0.19\textwidth]{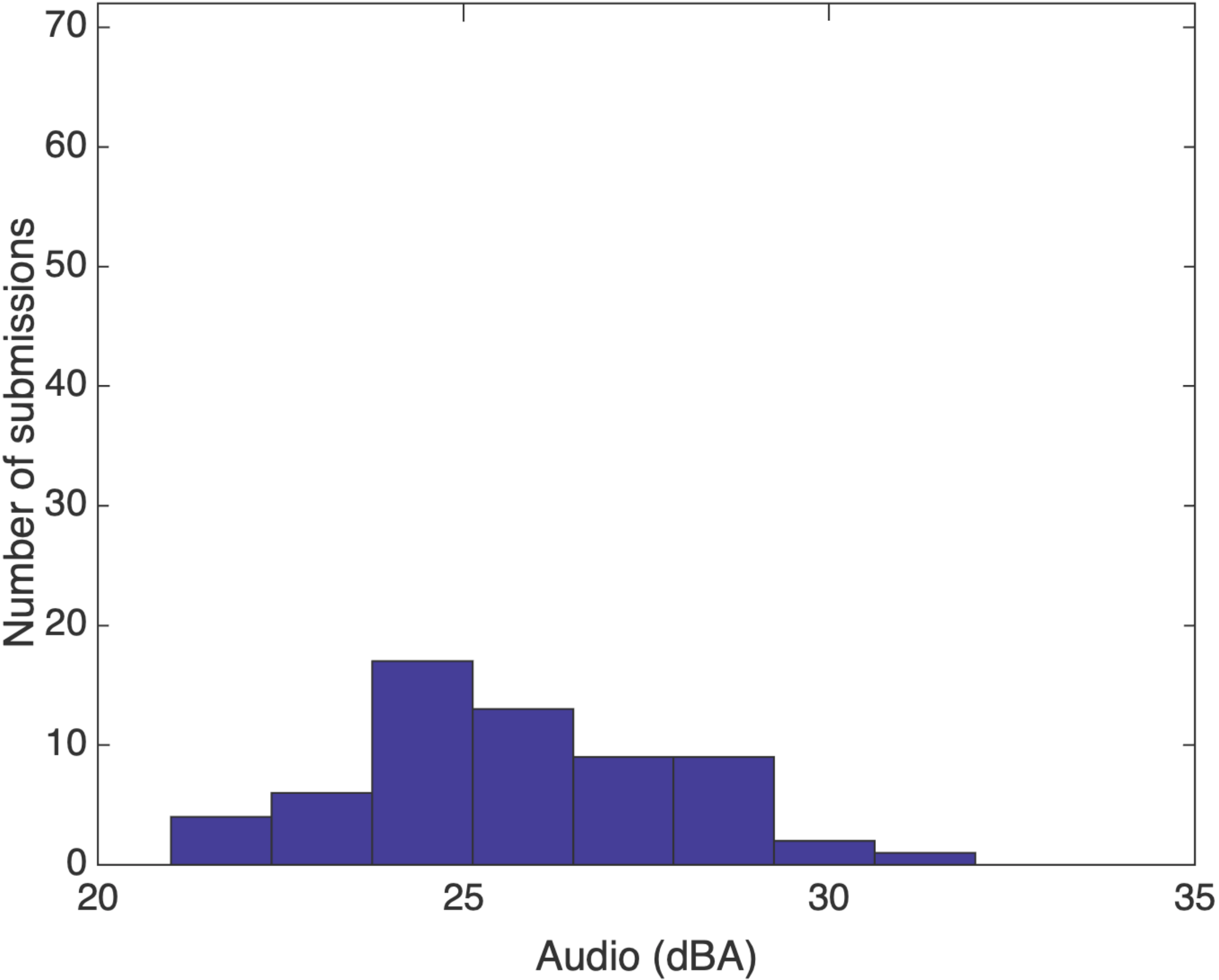}
    \label{fig:d5}}
    \caption{Distribution of sensor data for Experiment 1.}
    \label{secondA} 
\end{figure*}

\begin{figure*}[t!]
  \centering
  \captionsetup[subfloat]{farskip=0pt}%
\subfloat[Temperature]{\includegraphics[width=0.193\textwidth]{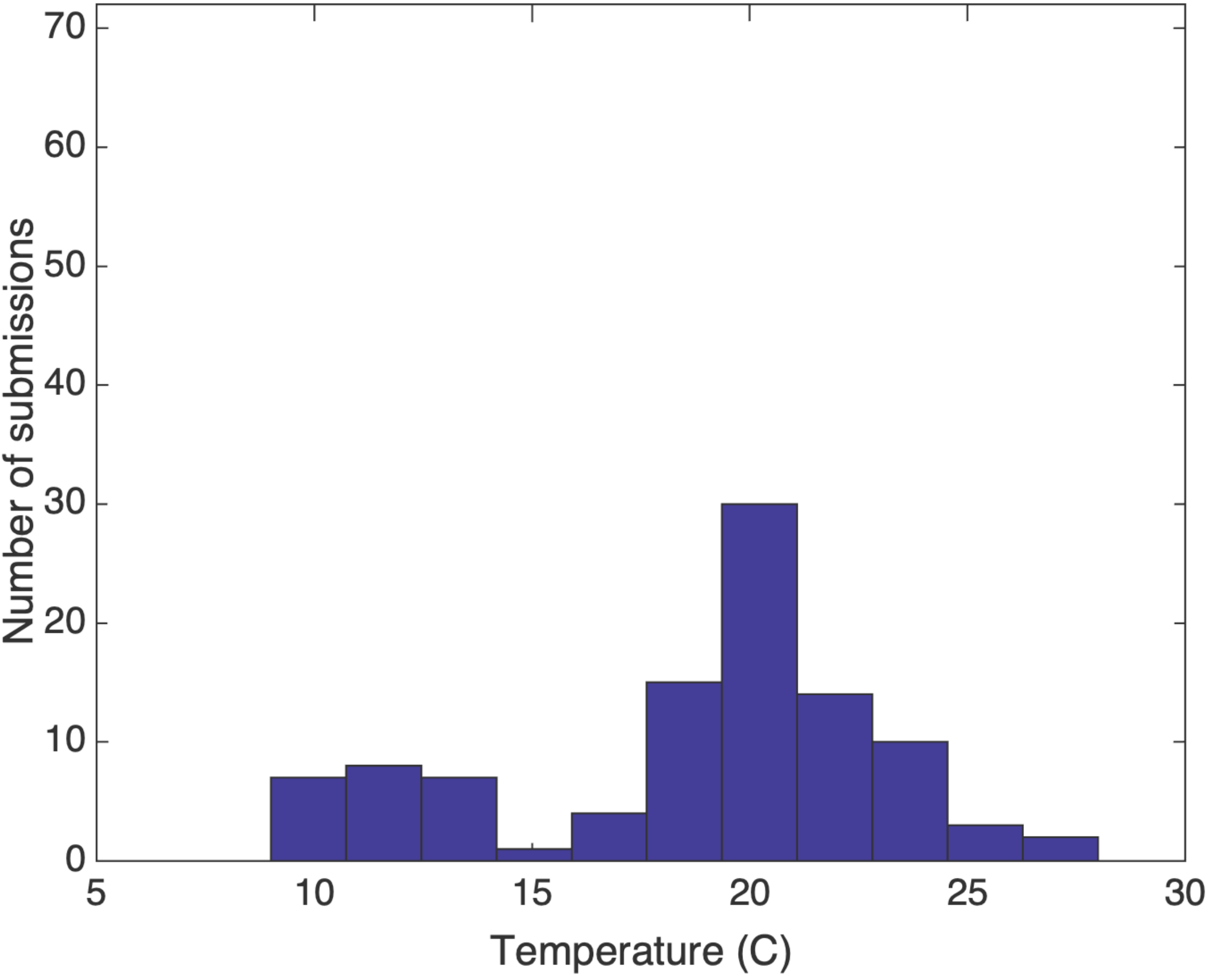} 
    \label{fig:d8}}\hfill
\subfloat[Humidity]{\includegraphics[width=0.19\textwidth]{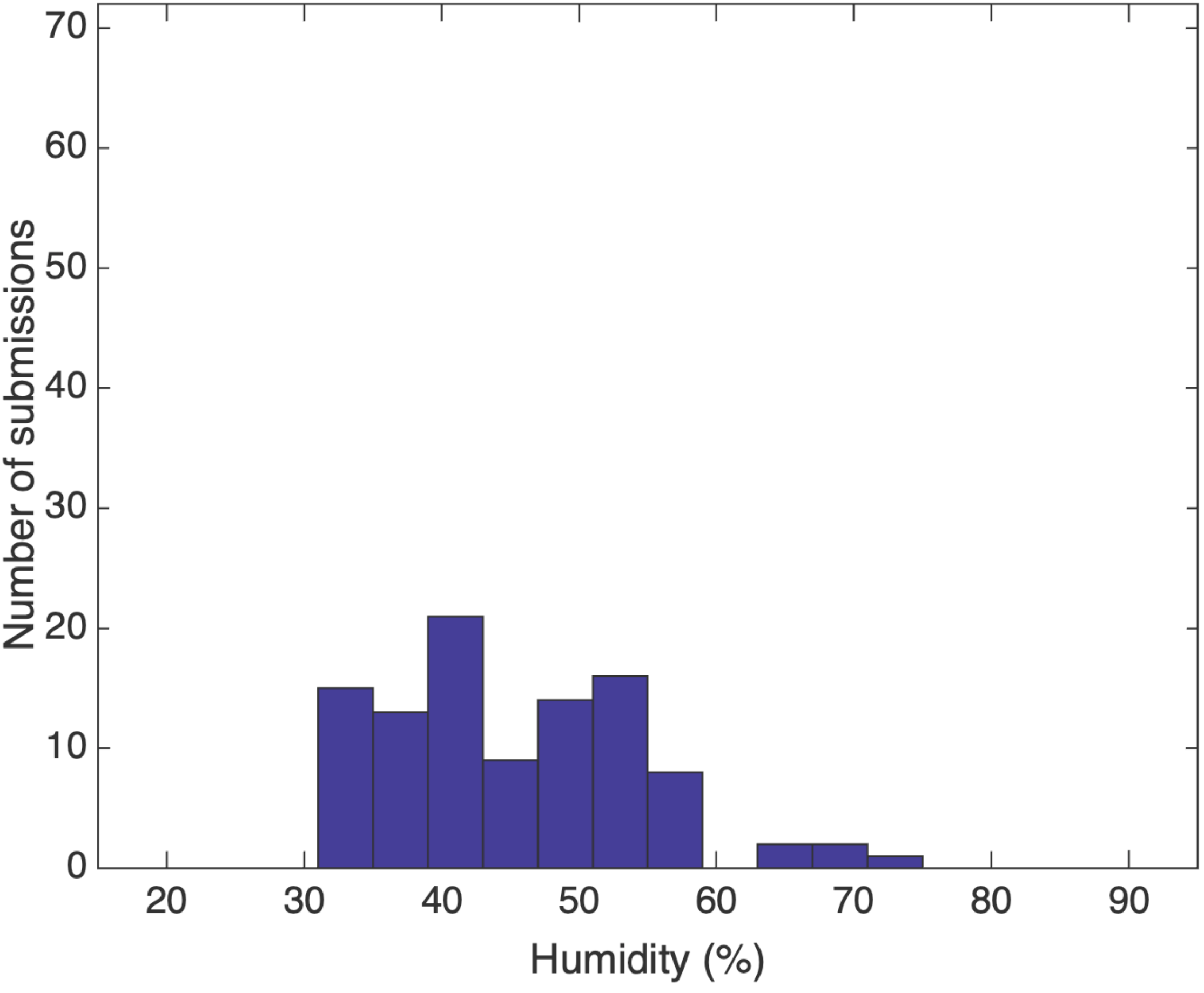}
    \label{fig:d2}}\hfill
\subfloat[Light]{\includegraphics[width=0.194\textwidth]{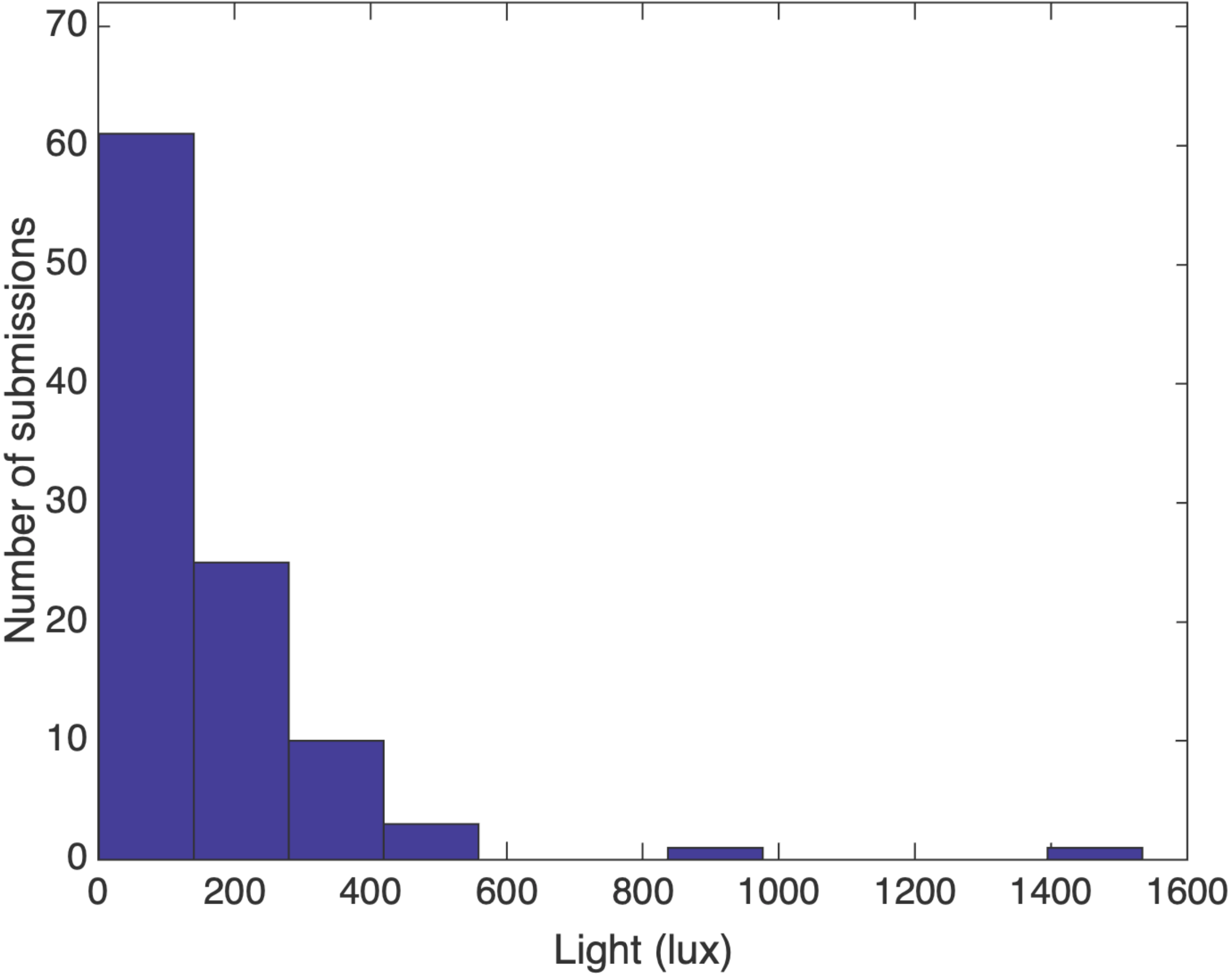}
       \label{fig:g1light}}\hfill
 \subfloat[Pressure]{\includegraphics[width=0.193\textwidth]{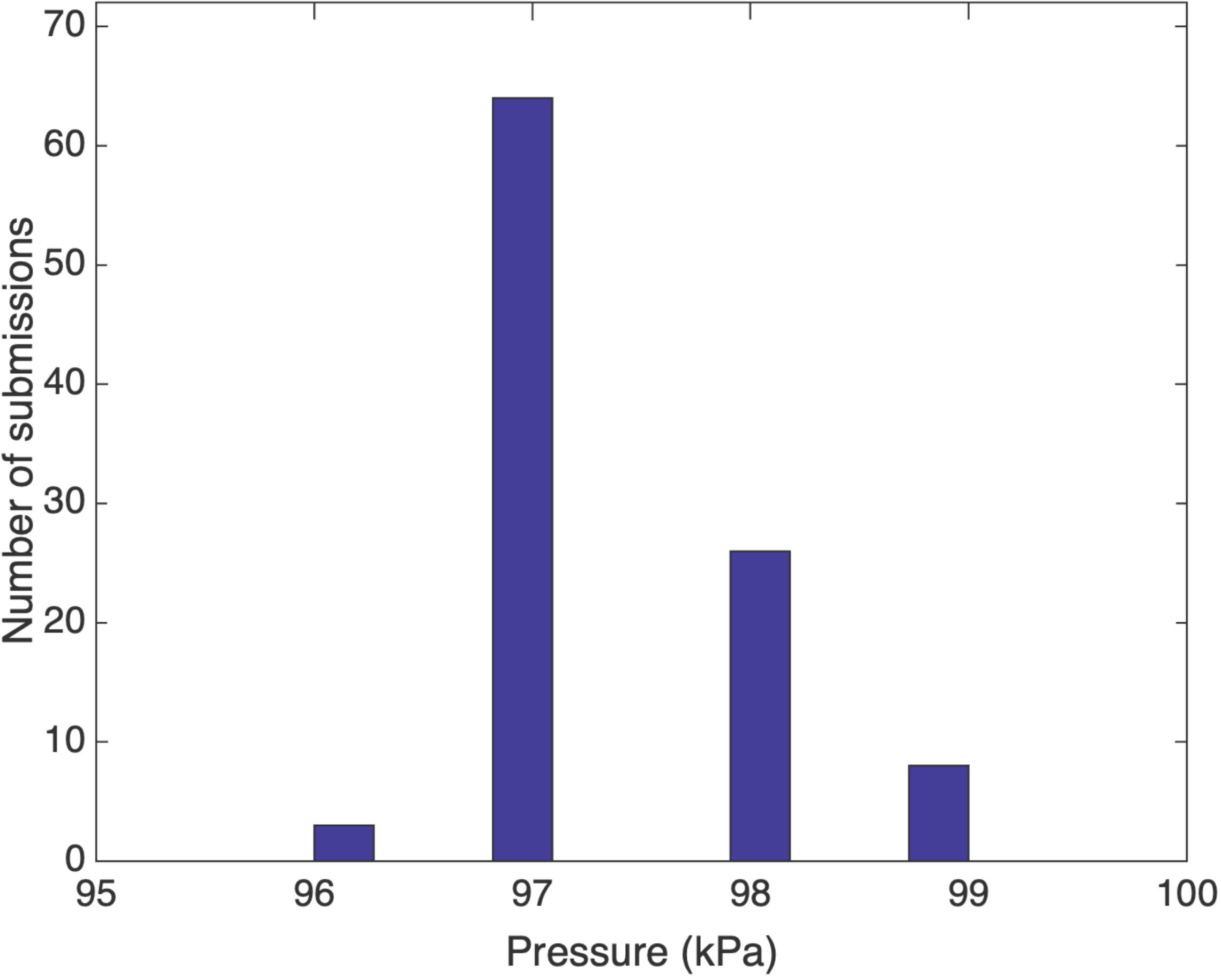}
    \label{fig:d4}}\hfill
\subfloat[Audio]{\includegraphics[width=0.193\textwidth]{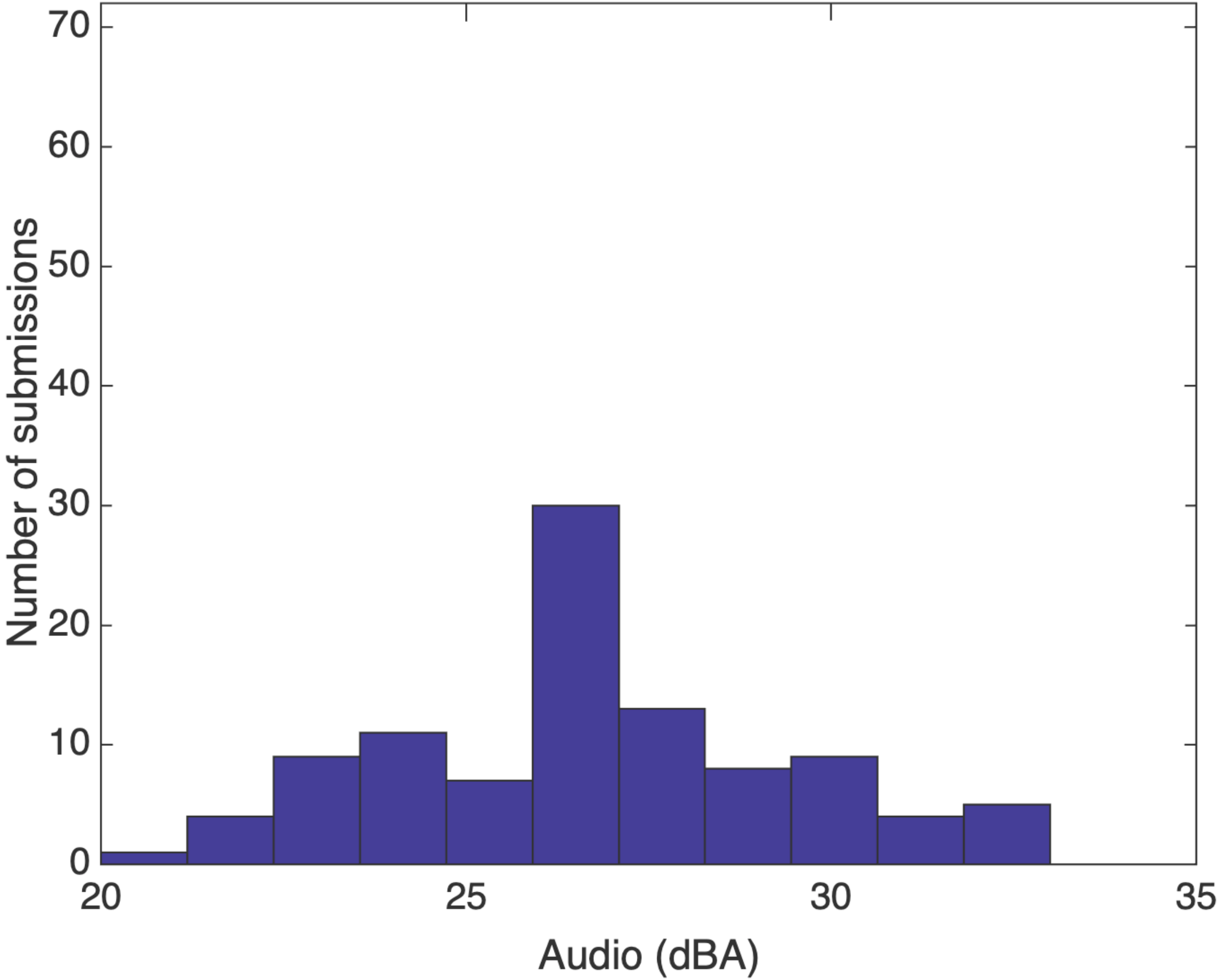}
    \label{fig:g1audio}}
    \caption{Distribution of sensor data for Experiment 2 - Group~A.}
    \label{secondB}
\end{figure*} 

\begin{figure*}[t!]
  \centering
  \captionsetup[subfloat]{farskip=0pt}%
\subfloat[Temperature]{\includegraphics[width=0.193\textwidth]{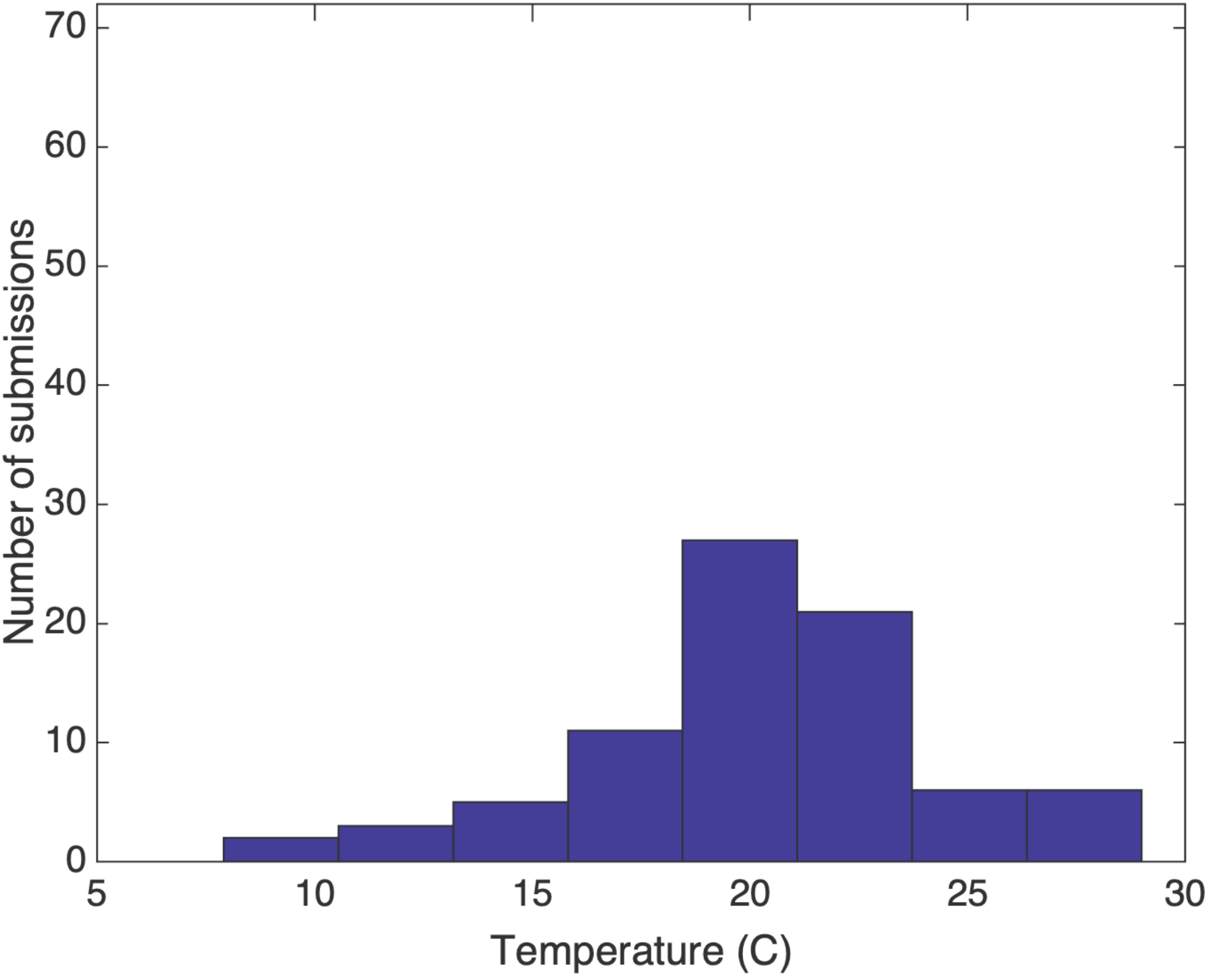}
    \label{fig:d6}}\hfill
\subfloat[Humidity]{\includegraphics[width=0.19\textwidth]{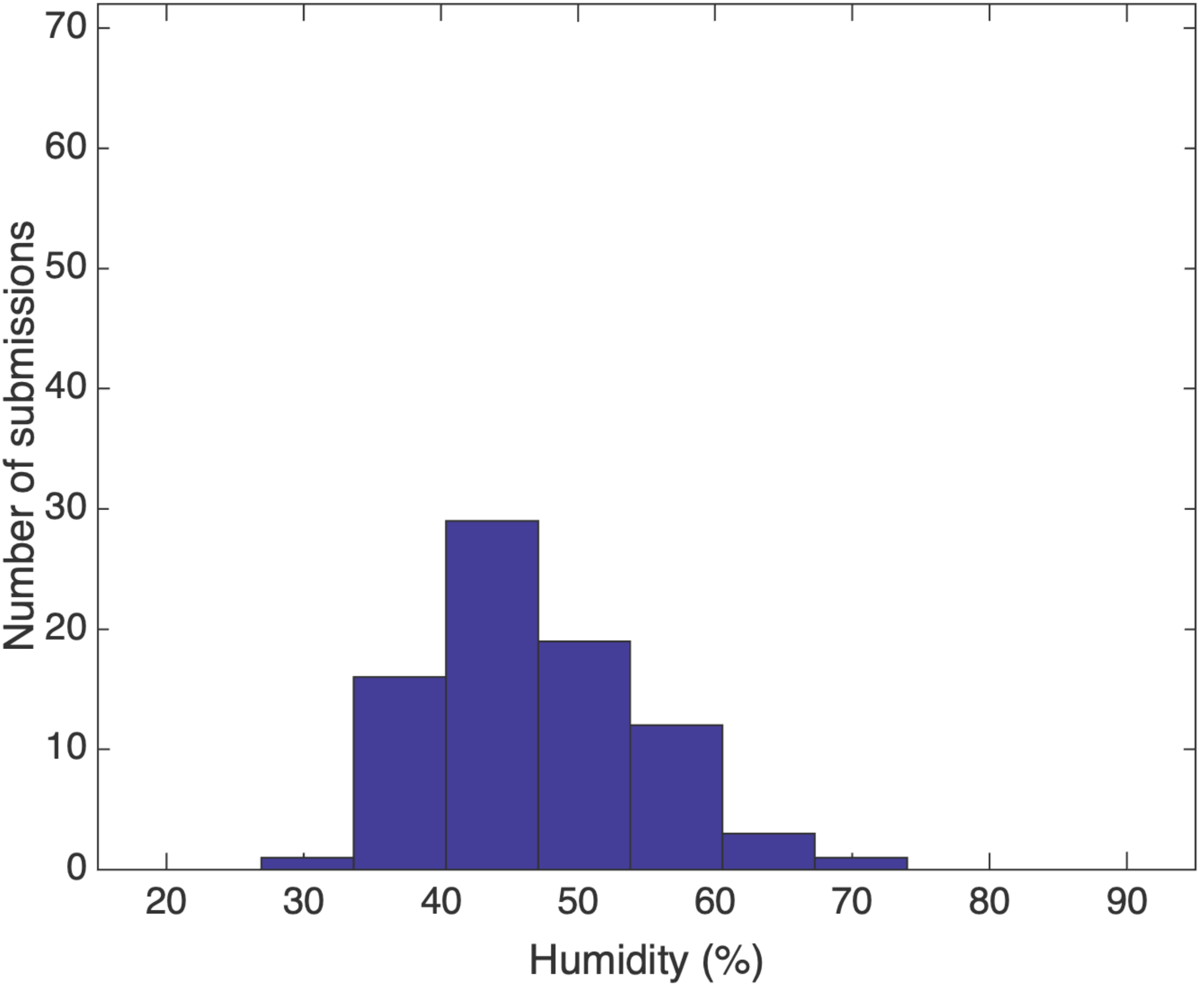}
    \label{fig:d7}}\hfill
\subfloat[Light]{\includegraphics[width=0.195\textwidth]{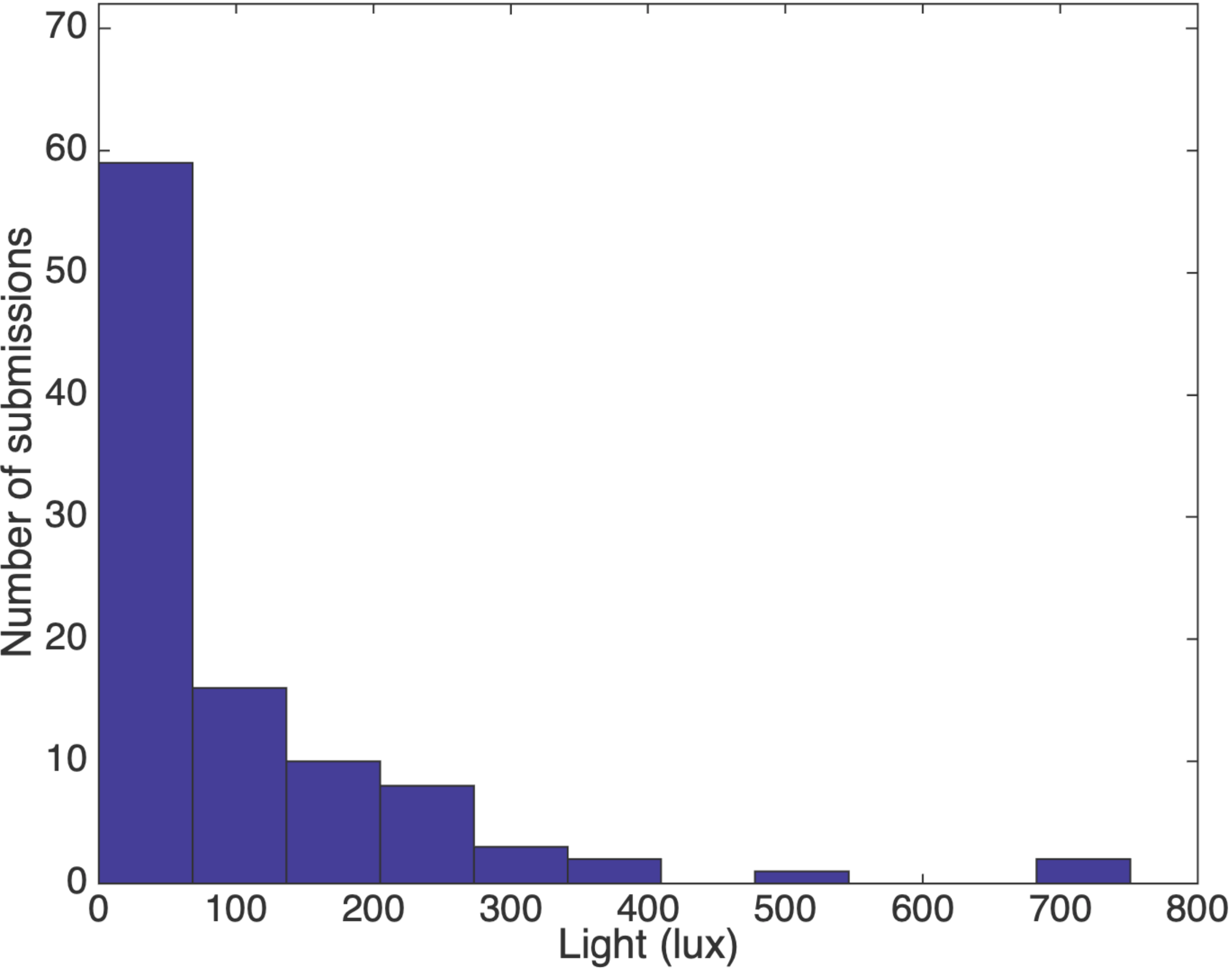}
       \label{fig:g2light}}\hfill
 \subfloat[Pressure]{\includegraphics[width=0.197\textwidth]{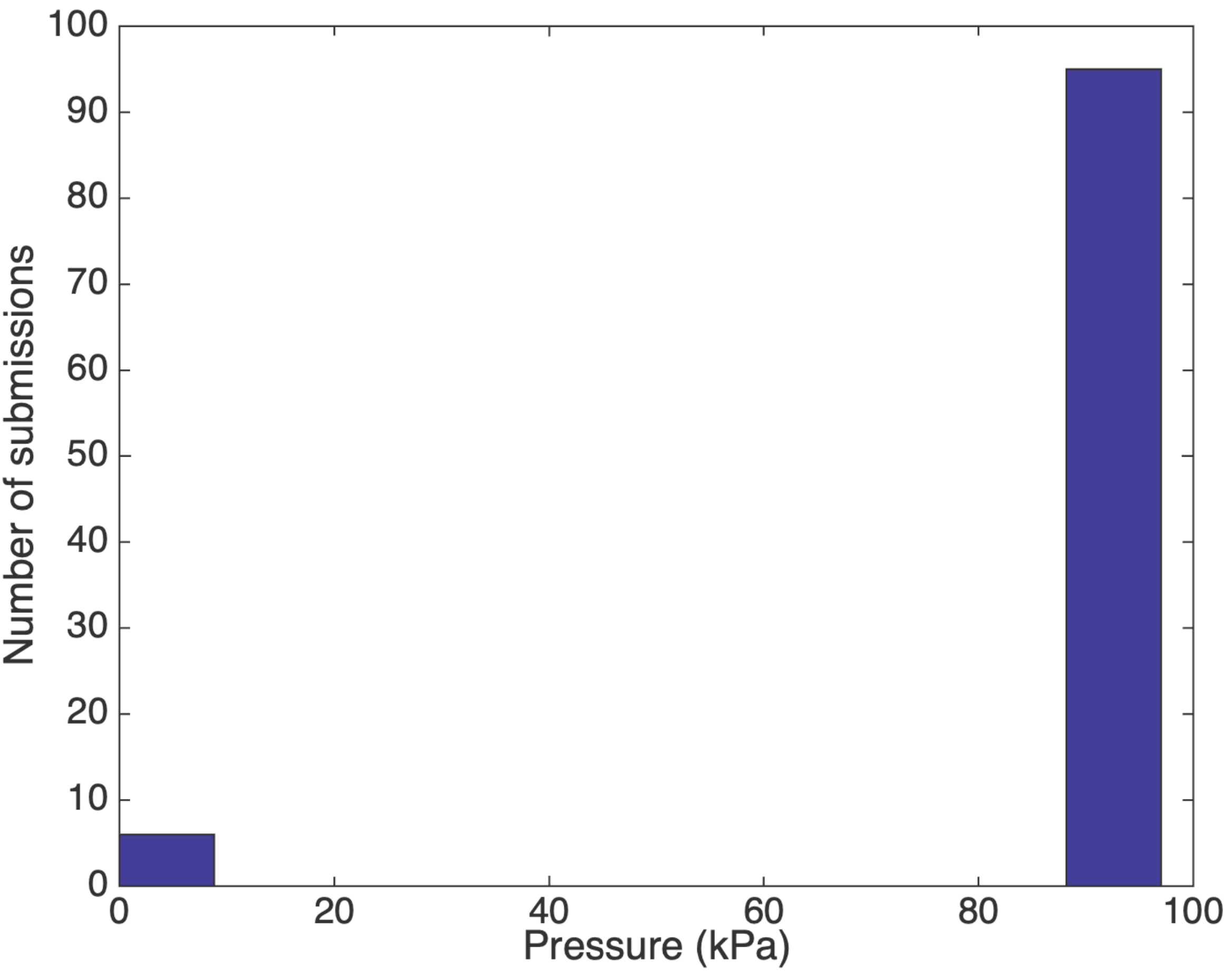}
    \label{fig:d4}}\hfill
\subfloat[Audio]{\includegraphics[width=0.19\textwidth]{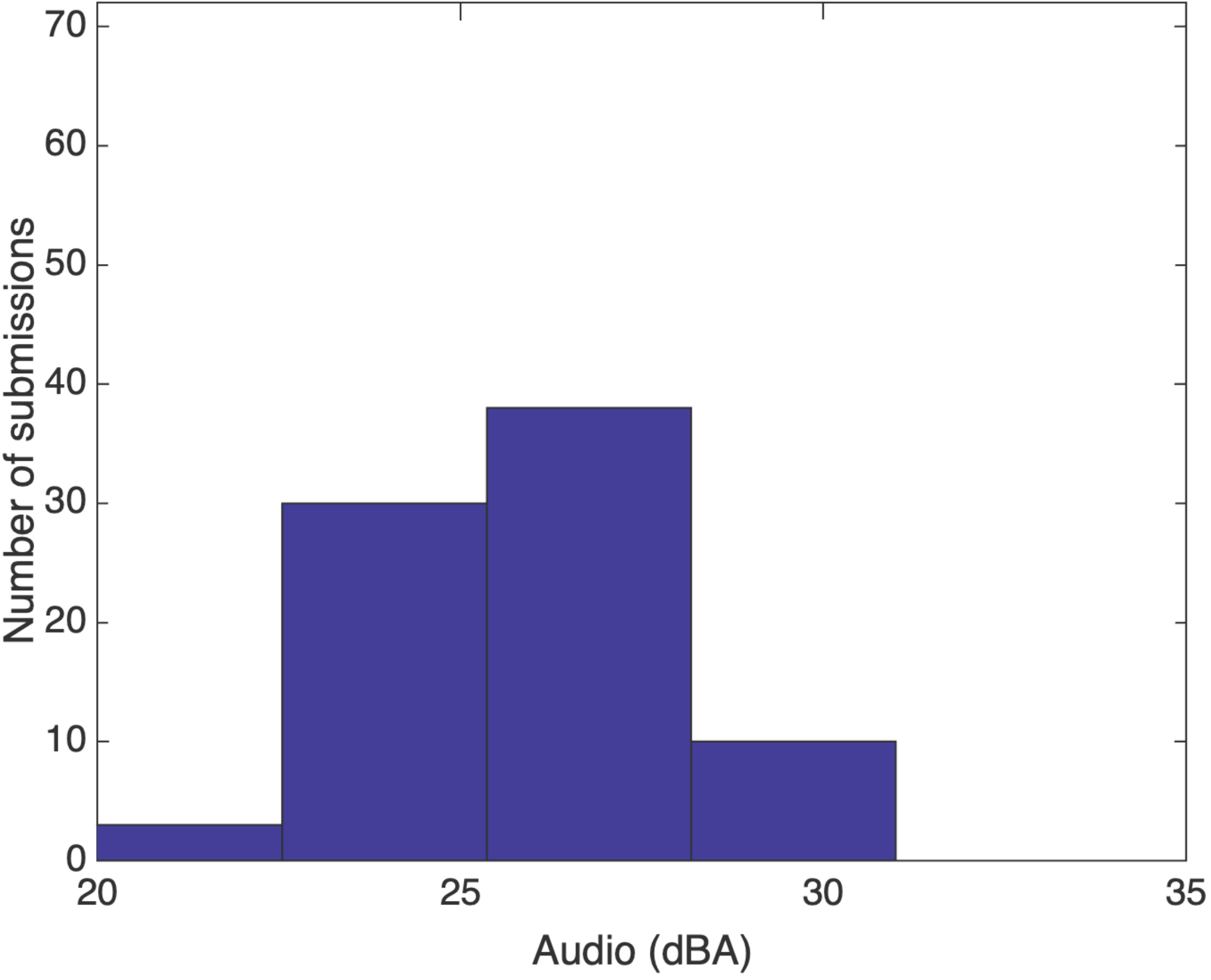}
    \label{fig:g2audio}}
    \caption{Distribution of sensor data for  Experiment 2 - Group~B.}
    \label{secondC}
\end{figure*}

 \begin{figure*}[t!]
  \centering
  \captionsetup[subfloat]{farskip=0pt}%
\subfloat[Temperature]{\includegraphics[width=0.193\textwidth]{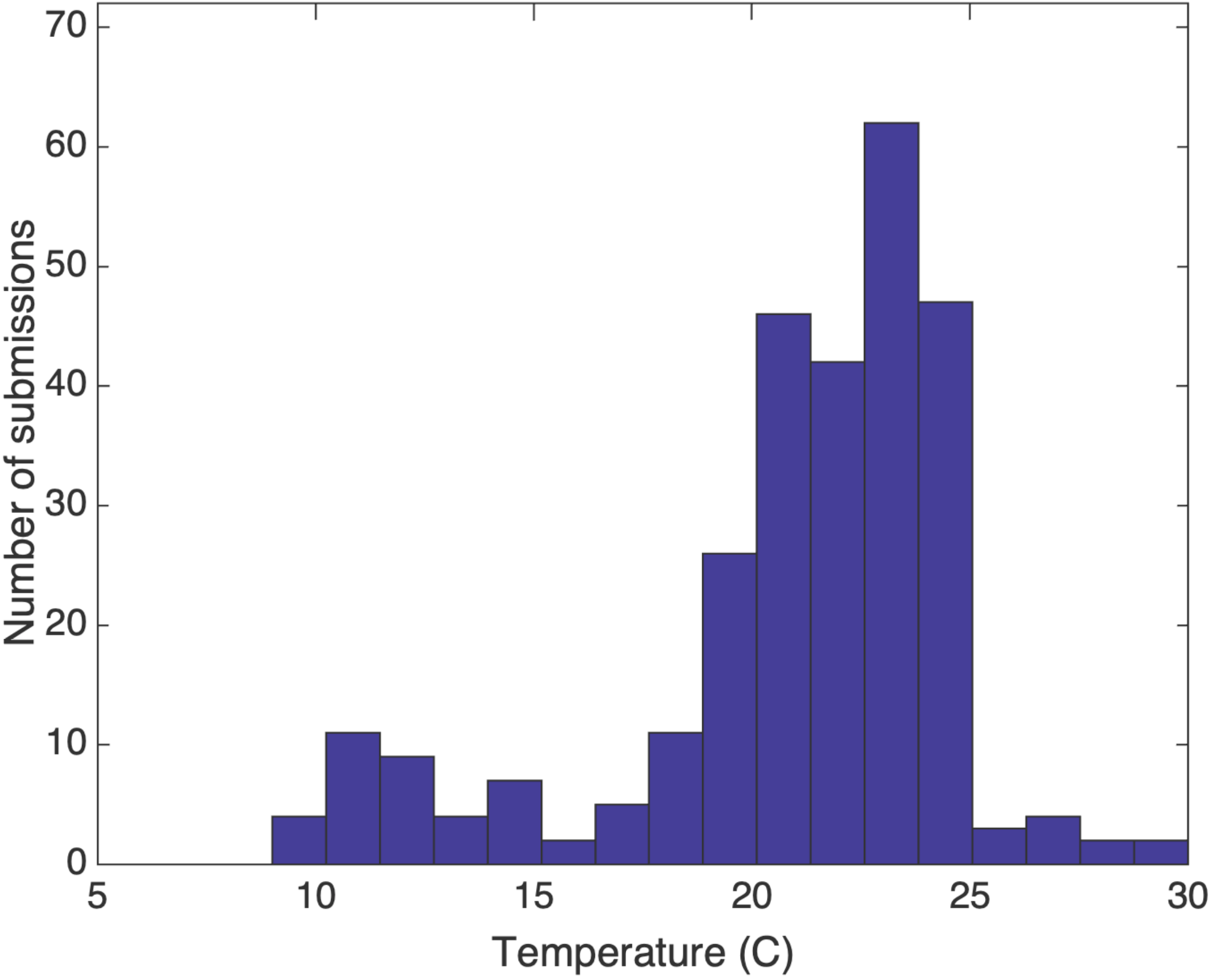}
    \label{fig:ex3-1}}\hfill
\subfloat[Humidity]{\includegraphics[width=0.19\textwidth]{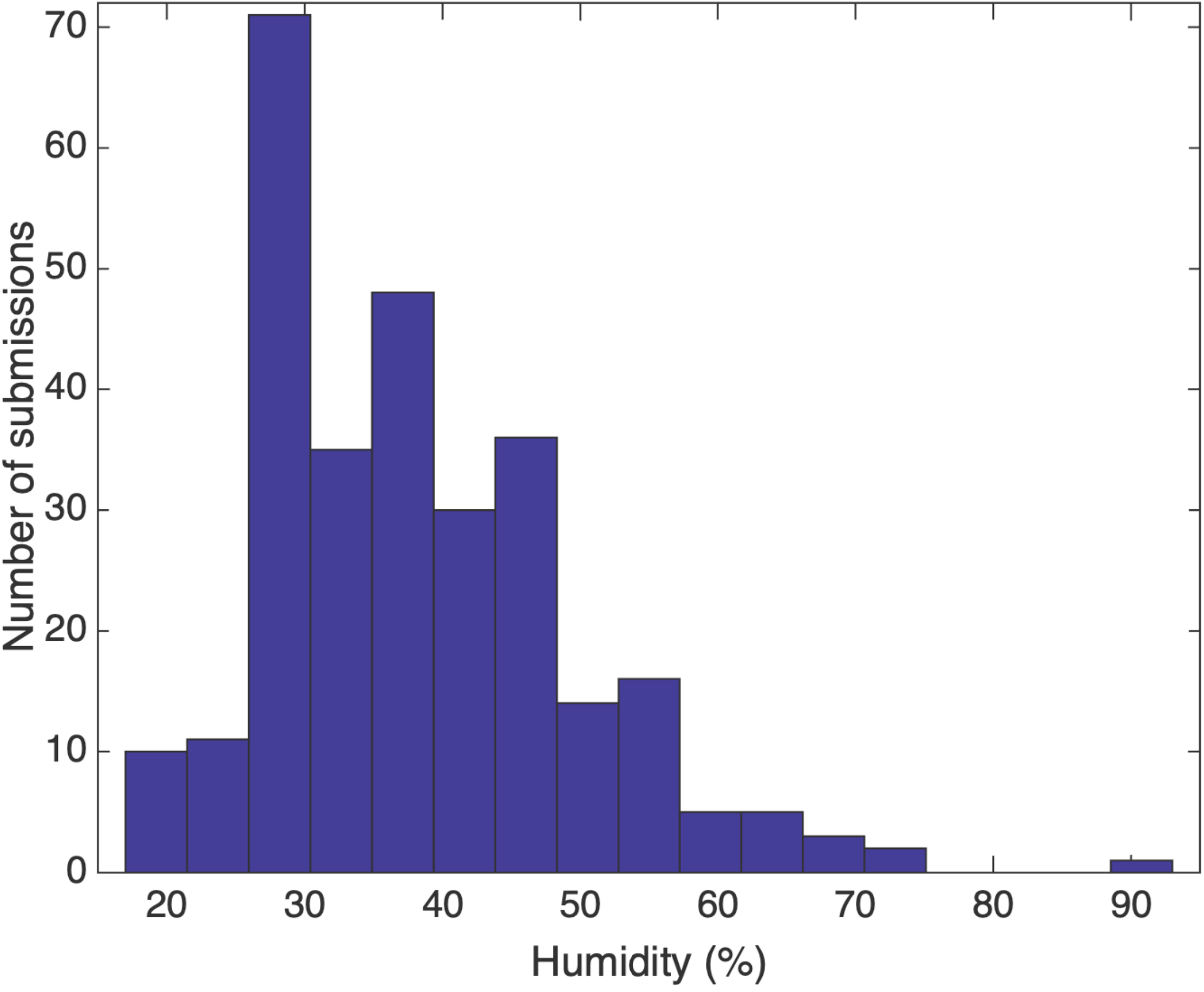}
    \label{fig:d7}}\hfill
\subfloat[Light]{\includegraphics[width=0.193\textwidth]{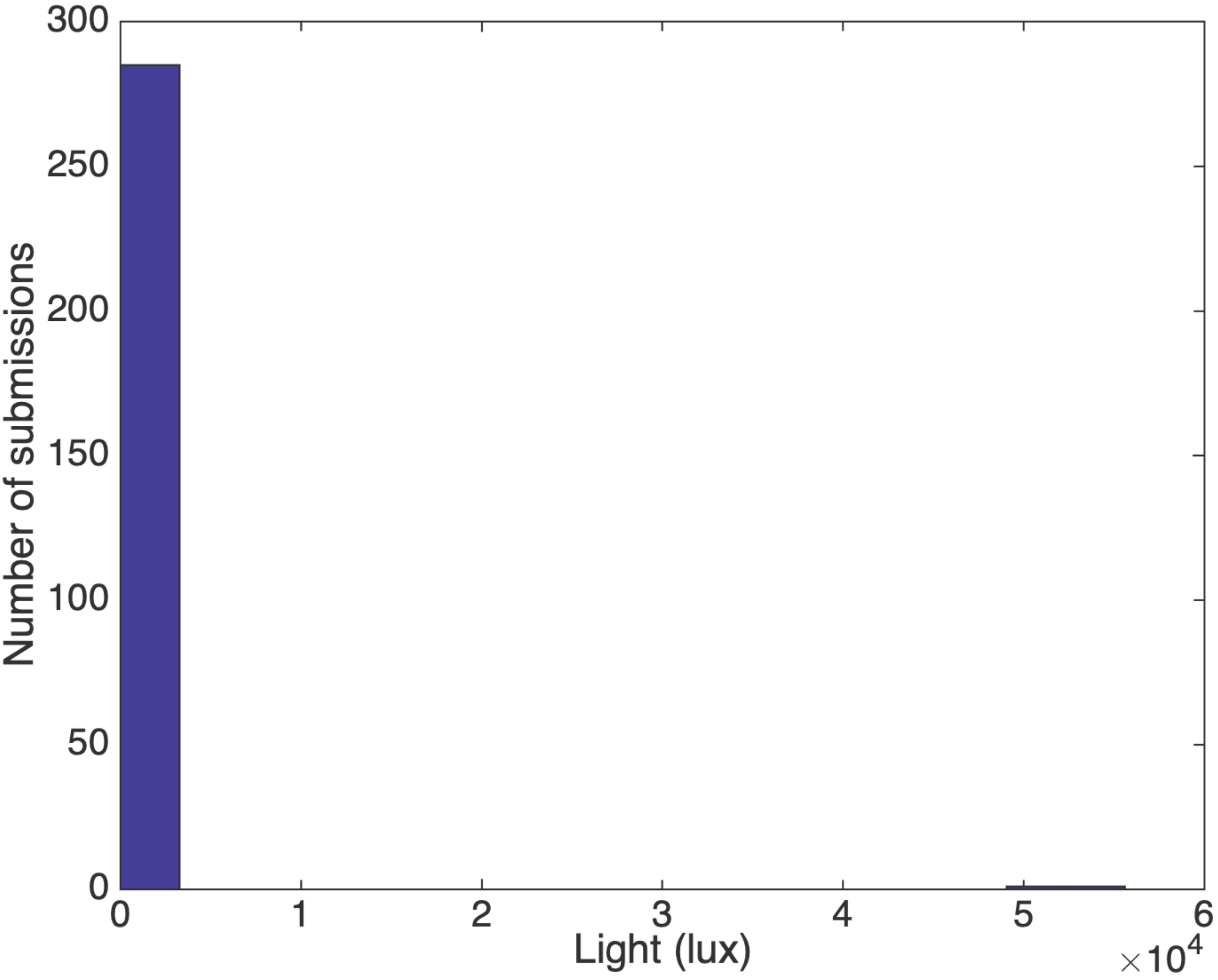}
       \label{fig:g3light}}\hfill
 \subfloat[Pressure]{\includegraphics[width=0.193\textwidth]{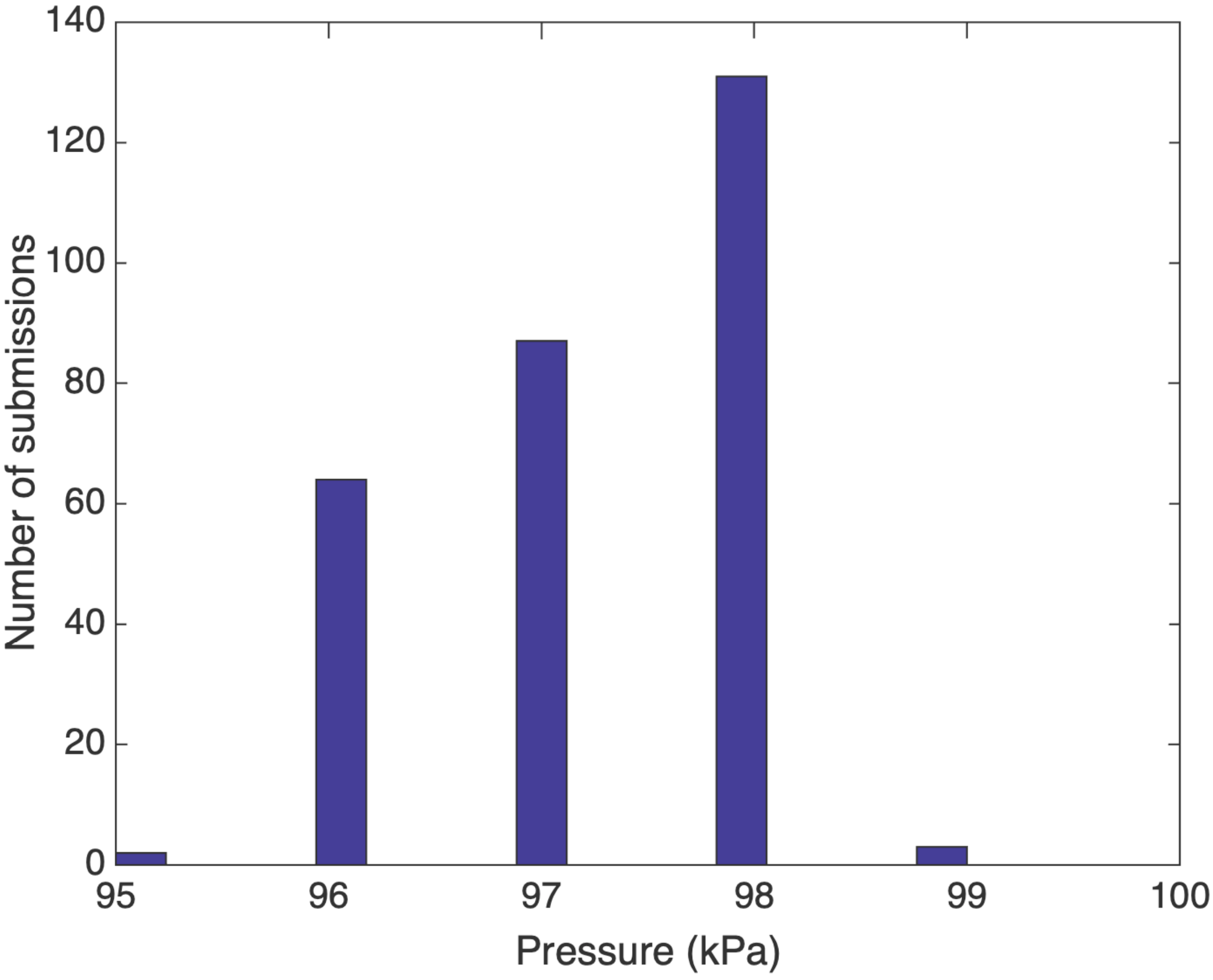}
    \label{fig:d4}}\hfill
\subfloat[Audio]{\includegraphics[width=0.19\textwidth]{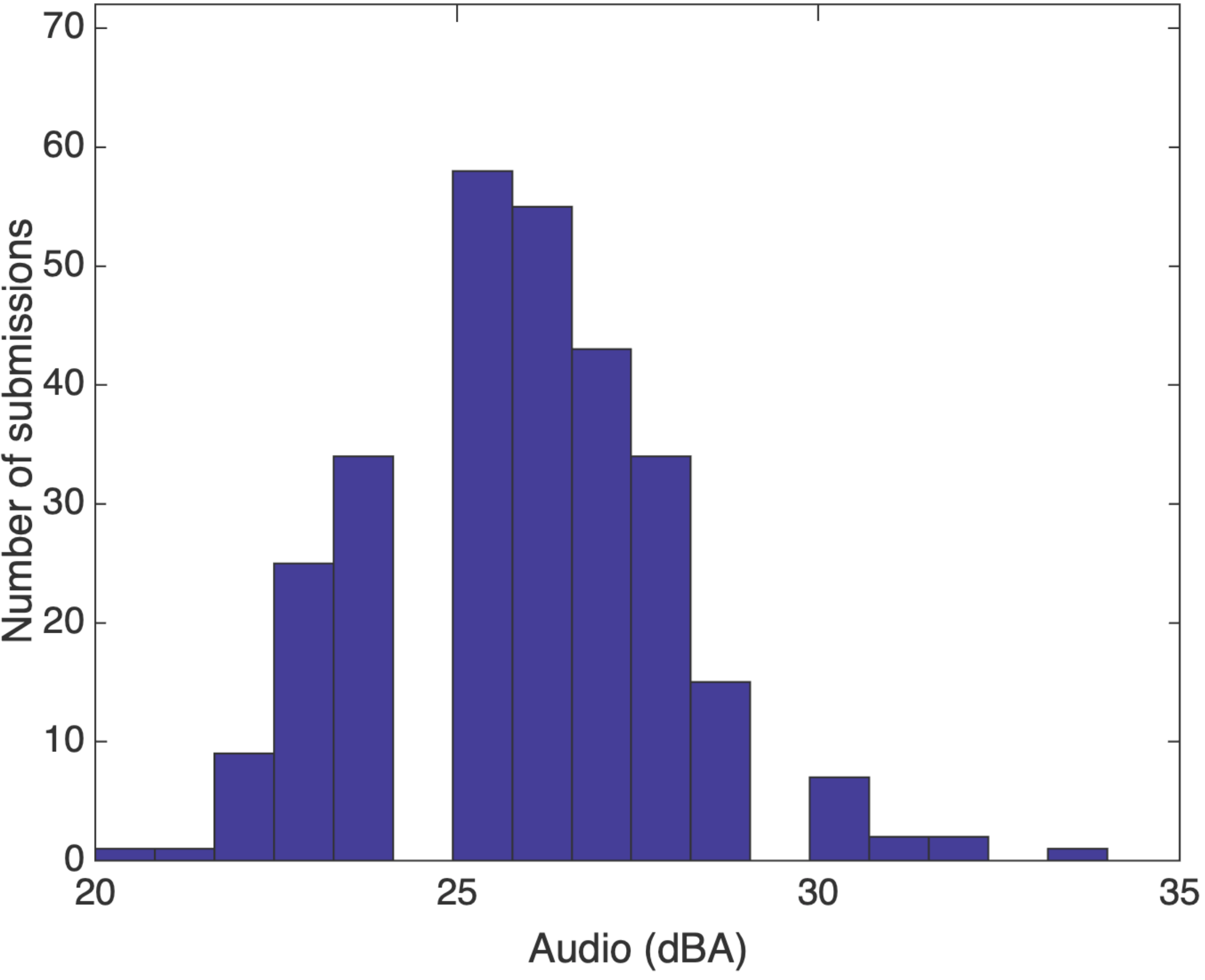}
    \label{fig:ex3-2}}
    \caption{Distribution of sensor data for Experiment 3.}
    \label{third}
\end{figure*}

  Following the training session, the participant completed the survey three times per day as instructed for a period of five days.  Following the experimental period, the participants return the loaned materials.  At this point, the data is analyzed and a correlation analysis is performed.

\subsection{Distribution of sensor data}
For the first experiment, 8 participants were selected to provide results over a period of 5 days. The participants included three men and five women, all of whom are University students. There were two major updates made since the initial testing phase.  The first update was replacing the movement variable, which was included in the initial testing~\cite{globalsip}, with audio.  Audio is a more significant variable and does not present the problem of affecting data from other sensors by having a SensorTag in motion.  The second update is to separate the survey results by survey type, rather than by topic of the question.  The results for the sensor data and survey results were otherwise calculated the same way as they were in the initial testing phase.  For each submission in each phase, the data for each variable were averaged.  For the survey results, an integer score was achieved for each of the three psychological surveys by incrementing by one for a positive answer.  Similar to~\cite{globalsip}, the results were parsed and checked for outliers.   This left valid data from 7 participants.  The distribution of each environmental variable is shown in Fig.~\ref{secondA}.

For the second experiment, 20 participants were recruited.  The participants were five women and fifteen men, all University students.  The participants were divided randomly into two groups of 10.  For the first week of the experiment, Group~A was asked to provide results over 5 days.  The following week, Group~B took their turn.  Each group then had access to the materials for a second week, resulting in up to 10 days of experimental data from each of our 20 participants.   The reason for the one week break between possession of materials was to prevent the participants from answering survey questions mindlessly and avoid memorizing the question order.   Of the submissions received to the server, 101 were valid in Group~A and 81 were valid from Group~B.  The distribution of each environmental variable for Group~A and for Group~B is seen in Fig.~\ref{secondB} and Fig.~\ref{secondC}, respectively. Again, we manage to have an acceptable distribution over different conditions.

For the third experiment, we recruited 34 participants and collected 287 valid submissions following the omission of outliers for faulty sensor data.  The distribution of each environmental variable for these submissions is shown in Fig.~\ref{third}.

\subsection{Methods of correlation}

The correlation coefficient, $r$, is a measure from $-1$ to $1$ which has both magnitude and direction~\cite{corrinfo}.  A coefficient of~$0$ indicates no association, and the association gets stronger as it approaches magnitude of~$1$.  A coefficient of~$1$ would indicate a perfect linear relationship.  However, this is unlikely in an experimental setting.  As data points diverge from the line indicating a linear relationship, the correlation coefficient decreases.  Additionally, the sign does not indicate the strength of a relationship, but rather whether it is a ``direct" or ``inverse" association.  The strength of a correlation coefficient can be roughly categorized as in Table~\ref{corrc}~\cite{corrinfo}.

\begin{table}[t!] 
\centering
\caption{Categorizing correlation coefficients ($r$ value)~\cite{corrinfo}.}
\label{corrc}
\begin{tabular}{ |l|r| }
  \hline
 \textbf{Weak or low}  & $\leq$ 0.35\\
 \hline
 \textbf{Moderate or modest} & 0.36 - 0.67\\
 \hline
 \textbf{Strong} &  0.68 - 1.0\\
 \hline
 \textbf{Very Strong}   & $\geq$ 0.9\\
 \hline
\end{tabular}%
\end{table}

It is possible to achieve a non-zero $r$ value for a relationship in which no correlation exists.  Therefore, it is important to analyze the significance of these values.  This can be done by calculating a $p$-value in addition to the $r$.  The $p$-value considers the chances of observing the given $r$ at random in a case that no correlation actually exists.  The $p$-value takes into account the sample size.  A small $p$-value indicates a statistically significant $r$, which allows for the rejection of the null hypothesis.  In the case of this experiment, the null hypothesis states ``there is no correlation between this environmental variable a survey result".  For the purposes of this experiment, the $p$-value must be lower than the standard $0.05$~\cite{corrinfo2} to be considered statistically significant and highlighted in the results to be discussed.

Three methods of correlation are used:
\begin{enumerate}[i.]
\item Pearson. Pearson's Product Moment correlation coefficient is used when both variables are normally distributed.  This coefficient is affected negatively by extreme values, which can exaggerate or minimize the degree of true association~\cite{pearsonmath}.

\item Spearman. Spearman Rank Coefficient is appropriate in similar cases to Pearson but accounts for the extreme values.  It is appropriate in cases where one or both variables are skewed or ordinal~\cite{spear}. It is important to use the Spearman coefficient when analyzing the light variable, as it is expected that most participants will take the survey within the indoor lighting range with exceptions where participants choose to take the survey outdoors.  Spearman will account for any extreme values.
\item Kendall. Kendall Rank Coefficient evaluates the degree of similarity between two sets of ranks among a set of objects.  The calculation is made based in part on the number of inversions in the rankings as one goes down the list~\cite{stats}.

\end{enumerate}

Once all data has been received, the correlation analysis is performed.  The three methods of correlation coefficient described were selected based on their required variable conditions.  The purpose of measuring all three is to compare the results of calculating an association using different methods. Emphasis is on the Spearman and Pearson coefficients when discussing the results.  Specifically, Spearman's coefficient, which takes into account extreme values.  We are looking for moderate correlation values that are confirmed by both Spearman and Pearson.  The confirmation is important because Pearson's coefficient is negatively affected by extreme values, ie. values that disrupt normal distribution.  Spearman's coefficient accounts for these. 
 

\section{Results and Discussion}  \label{results}

The following subsections present and evaluate the results for all three experiments.

\subsection{Correlation analysis of survey responses}

Correlation coefficients were calculated for each correlation method between each of the three survey sections and the additional question concerning the number of people around.  For Experiment 1, the coefficients are presented in Table~\ref{corr1}.  The greatest Pearson correlation between different variables, $0.7277$, exists between the PSS and K10 surveys, indicating a correlation between stress and psychological distress. Additionally, Pearson correlation of $0.6187$ was determined between PSQI and K10, indicating a correlation between sleep quality and distress.  The stronger of the two correlations, PSS and K10, also present a Spearman correlation of $0.8456$ and a Kendall correlation of $0.7056$.  However, Spearman and Kendall do not show any correlations between PSQI and K10. 

\begin{table}[t!] 
\centering
\caption{Correlation coefficients between all survey results in Experiment 1.}
\label{corr1}
\begin{tabular}{ |p{1.5cm}|p{1.5cm}|p{1cm}|p{1cm}|p{1cm}| }
\hline
  & \# of People & PSQI & PSS & K10   \\
\hline
\hline
 \multicolumn{5}{|c|}{\textbf{Pearson correlation}} \\
  \hline\hline
 \# of People  &  &-0.0399 & 0.2094 & -0.0717\\
 PSQI&  -0.0399 &   & 0.3597 & \textbf{0.6187}\\
 PSS& 0.2094 & 0.3597 &  & \textbf{0.7277} \\
 K10   & -0.0717 & \textbf{0.6187} & \textbf{0.7277} &  \\
 \hline
 \hline
 \multicolumn{5}{|c|}{\textbf{Spearman correlation}} \\
  \hline\hline
 
 People  & &-0.0657 & 0.2109 & 0.0031\\
 PSQI&  -0.0657 &  & 0.2489 & 0.3978\\
 PSS& 0.2109 & 0.2489 &  & \textbf{0.8456} \\
 K10   & 0.0031 & 0.3978 & \textbf{0.8456} & \\
 \hline
 \hline
 \multicolumn{5}{|c|}{\textbf{Kendall correlation}} \\
  \hline\hline
 People  & &-0.0551 & 0.1737 & -0.0026\\
 PSQI&  -0.0551 &  & 0.1991 & 0.3130\\
 PSS& 0.1737 & 0.1991 &  & \textbf{0.7056} \\
 K10   & 0.0026 & 0.3130 & \textbf{0.7056} & \\
 \hline
\end{tabular}%
\end{table}

For Experiment 2, as shown in Table~\ref{corr4}, the survey results from Group~A show a strong Pearson $0.5854$ and Spearman $0.5408$ correlation between the PSS and K10 results.  In Group~B, shown in Table~\ref{corr71}, this same correlation was even stronger, with a  Pearson value of $0.6325$, a Spearman value of $0.6435$, and a Kendall value of $0.5213$.  This mirrors the results of the first experimental set, indicating correlation between stress and distress.  There were no other correlations of less than 0.5 in Experiment 2 between survey results.

\begin{table}[t!] 
\centering
\caption{Correlation coefficients between all survey results in Experiment 2 - Group~A.}
\label{corr4}
\begin{tabular}{ |p{1.5cm}|p{1.5cm}|p{1cm}|p{1cm}|p{1cm}| }
\hline
  & \# of People & PSQI & PSS & K10   \\
  \hline\hline
   \multicolumn{5}{|c|}{\textbf{Pearson correlation}} \\
 \hline\hline
 \# of People  & &-0.0511 & -0.2248 &-0.2221 \\
 PSQI& -0.0511  &  & 0.0966 & 0.2825\\
 PSS& -0.2248 & 0.0966 &  & \textbf{0.5854} \\
 K10   & -0.2221 & 0.2825& \textbf{0.5854} & \\
  \hline\hline
  \multicolumn{5}{|c|}{\textbf{Spearman correlation}} \\
 \hline\hline
 \# of People  & &-0.0436 & -0.2162 & -0.2647\\
 PSQI& -0.0436  &  & 0.0959 & 0.2666\\
 PSS& -0.2162 & 0.0959 &  & \textbf{0.5408} \\
 K10   & -0.2647 & 0.2666& \textbf{0.5408} & \\
 \hline\hline
   \multicolumn{5}{|c|}{\textbf{Kendall correlation}} \\
 \hline\hline
\# of People  & & -0.0374 & -0.1774 & -0.2214\\
 PSQI& -0.0374  &  & 0.0771 & 0.2204\\
 PSS& -0.1774 & 0.0771 &  & 0.4478 \\
 K10   & -0.2214 & 0.2214 & 0.4478 & \\
 \hline
\end{tabular}%
\end{table}

\begin{table}[t!] 
\centering
 \caption{Correlation coefficients between all survey results in the Experiment 2 - Group~B.}
\label{corr71}
\begin{tabular}{ |p{1.5cm}||p{1.5cm}|p{1cm}|p{1cm}|p{1cm}| }
\hline
  &\# of  People & PSQI & PSS & K10   \\
     \hline\hline
   \multicolumn{5}{|c|}{\textbf{Pearson correlation}} \\
 \hline\hline
 \# of People  & & -0.1085& -0.3723 & -0.4494\\
 PSQI&  -0.1085 &  & 0.1696 & 0.2734\\
 PSS& -0.3723 & 0.1696 &  & \textbf{0.6325} \\
 K10   & -0.4494 & 0.2734& \textbf{0.6325} & \\
    \hline\hline
   \multicolumn{5}{|c|}{\textbf{Spearman correlation}} \\
 \hline\hline
\# of  People  &  & -0.1757 & -0.3756 & -0.4451\\
 PSQI&  -0.1757 &  & 0.2320 & 0.3584\\
 PSS& -0.3756 & 0.2320 &  & \textbf{0.6435} \\
 K10   & -0.4451 & 0.3584& \textbf{0.6435} & \\
  \hline\hline
   \multicolumn{5}{|c|}{\textbf{Kendall correlation}} \\
 \hline\hline
 \# of People  & & -0.1472 & -0.3004 & -0.3844\\
 PSQI& -0.1472  &  & 0.1894 & 0.2737\\
 PSS& -0.3004 & 0.1894 &  & \textbf{0.5213} \\
 K10   & -0.3844 & 0.2737& \textbf{0.5213} & \\
 \hline
 \end{tabular}
\end{table}

Table~\ref{3corr1} presents the correlation analysis for Experiment 3.  The analysis shows moderate positive Pearson and Spearman correlation between PSQI and PSS, PSQI and K10, as well as PSS and K10.  The strongest relationship $0.6392$ in Pearson exists between the PSS and K10 surveys.  This is encouraging as a repeated result as it shows an expected relationship between stress and distress as measured by our surveys.

\begin{table}[t!] 
\centering
 \caption{Correlation coefficients between all survey results in the Experiment 3.}
\label{3corr1}
\begin{tabular}{ |p{1.5cm}||p{1.5cm}|p{1cm}|p{1cm}|p{1cm}| }
\hline
  &\# of  People & PSQI & PSS & K10   \\
     \hline\hline
   \multicolumn{5}{|c|}{\textbf{Pearson correlation}} \\
 \hline\hline
 \# of People  & & -0.0622 & -0.1355  &-0.3245 \\
 PSQI& -0.0622   &  &\textbf{0.4460}  &\textbf{0.4009} \\
 PSS&  -0.1355& \textbf{0.4460}&  & \textbf{0.6392} \\
 K10   & -0.3245  & \textbf{0.4009}&\textbf{0.6392}  & \\
    \hline\hline
   \multicolumn{5}{|c|}{\textbf{Spearman correlation}} \\
 \hline\hline
\# of  People  &  & -0.1897 & -0.3263 & -0.2411\\
 PSQI&  -0.1897 &  & \textbf{0.4481}&  \textbf{0.4146}\\
 PSS& -0.3263 & \textbf{0.4481}&  &  \textbf{0.5961} \\
 K10   & -0.2411 & \textbf{0.4146}& \textbf{0.5961} & \\
  \hline\hline
   \multicolumn{5}{|c|}{\textbf{Kendall correlation}} \\
 \hline\hline
 \# of People  & & -0.1451 & -0.1833 & -0.2814\\
 PSQI& -0.1451  &  & 0.3554 & 0.3469\\
 PSS& -0.1833 & 0.3554 &  & \textbf{0.4779} \\
 K10   & -0.2814 &  0.3469& \textbf{0.4779} & \\
 \hline
 \end{tabular}
\end{table}

\subsection{Correlation analysis of survey responses to environmental variables} 

Correlation between the survey results and the environmental data is also examined.  For Experiment 1, as shown in Table~\ref{ex1temp}, the strongest Pearson correlation between PSS and light, $-0.5641$, indicating a negative correlation between increased stress and increased luminosity.  This correlation is presented in Spearman as well, with a value of $-0.5753$, but not in Kendall.  In fact, Kendall coefficients determined no correlation between our surveys and environmental variables.  However, the normal distribution of our five variables observed in Fig.~\ref{secondA} means that Pearson correlation is valid, as is Spearman due to the ordinal nature of the variables.  Pearson found that three of our variables are correlated with the PSS scale: temperature, humidity, and light.  Spearman confirms the correlation of PSS with temperature and light.  It further defines correlation between K10 responses and temperature, humidity, and light.   All moderate correlations discussed, were confirmed as significant, when $p$-value is less than 0.05.  The low $p$-values indicate that even the lower correlation coefficients are statistically significant.

\begin{table*}[t!] 
\centering
\caption{Correlation between variables and surveys for Experiment 1.}
\label{ex1temp}
\begin{tabular}{ |c|c|c|c|c|c|c|c|c|}
 \hline
  & \multicolumn{2}{|c|}{\# of People} & \multicolumn{2}{|c|}{PSQI} & \multicolumn{2}{|c|}{PSS} & \multicolumn{2}{|c|}{K10}   \\
 \hline
 & $r$ & $p$-value & $r$ & $p$-value & $r$ & $p$-value & $r$ & $p$-value  \\ \hline
   \hline
\multicolumn{9}{|c|}{\textbf{Pearson Correlation}}     \\
  \hline
 
 Temperature  & 0.1035     & 0.1777   & 0.3461  & 0.1901  & 0.5095 &0.0000253          &  0.49       & 0.0012\\
 Pressure        &   -0.1971  &  0.5097 &-0.1713   & 0.6443    & -0.0967 & 0.9303            &-0.0964           &0.9642 \\
 Humidity         & 0.0832     &  0.8564   &-0.1976     &0.1392     & -0.5128 &0.0004418 &-0.4453            &0.0017\\
 Light               &-0.0613     &  0.618   & -0.1244    & 0.9545    & \textbf{-0.5641} &0.000006  &-0.3769 &0.0008575 \\
 Audio             &  0.2594     & 0.2333  &-0.0687   &0.3623     & 0.2614 & 0.096             &0.1185          &0.1477\\
\hline \hline
\multicolumn{9}{|c|}{\textbf{Spearman Correlation}}     \\
  \hline
 Temperature  & 0.0802	& 0.6428 	& 0.3093	&0.0167 	& 0.5063 	&0.0000856	& 0.5389		&0.000029    \\
 Pressure       &  -0.1671	& 0.1656  	& -0.1861	&0.1934  	& 0.0415 	&0.8	& -0.0078		& 0.9412    \\
 Humidity        & 0.1093	& 0.4942	& -0.2253	&0.0686	& -0.4604 	&0.0002637	&-0.5069 		&0.0000689 \\
 Light              & -0.0173	& 0.8136	& -0.0358 	&0.2998	& \textbf{-0.5753}	&0.0000017	& -0.5381&0.0000073\\
 Audio            &  0.2612	&  0.0473 	& -0.1507	& 0.7777 	& 0.259	&0.0392	&0.0975		&  0.3657\\
 \hline
 \hline
\multicolumn{9}{|c|}{\textbf{Kendall Correlation}}     \\
\hline 
 Temperature  	& 0.0653	& 0.6545	& 0.238	& 0.019	& 0.397	&0.0001093 	&0.4375				& 0.0000398 \\
 Pressure		&  -0.1573	&  0.1635	& -0.1575	& 0.1884	& 0.0273	&0.8753  	&-0.0051				&  0.9152   \\
 Humidity		& 0.0888	& 0.4955 	& -0.1796	&0.0651	& -0.3315	& 0.0005896	& -0.3871			&0.0001365 \\
 Light   		& -0.0138	& 0.8131	& -0.0304	&0.2874	&-0.4179	&0.0000081 	& -0.4114		& 0.0000184\\
 Audio		&  0.2194	& 0.0494 	& -0.1163	& 0.7818	& 0.181	&0.0569& 0.0792			& 0.3502\\
\hline
\end{tabular}%

\end{table*}

\begin{table*}[t!] 
\centering
\caption{Correlation between variables and surveys for Experiment 2 - Group~A.}
\label{ex2tempA}
\begin{tabular}{  |c|c|c|c|c|c|c|c|c|}
 \hline
  & \multicolumn{2}{|c|}{\# of People} & \multicolumn{2}{|c|}{PSQI} & \multicolumn{2}{|c|}{PSS} & \multicolumn{2}{|c|}{K10}   \\
 \hline
 & $r$ & $p$-value & $r$ & $p$-value & $r$ & $p$-value & $r$ & $p$-value  \\ \hline
   \hline
\multicolumn{9}{|c|}{\textbf{Pearson Correlation}}     \\
  \hline

 Temperature  & 0.3798 & 0.001649 &  -0.0985 &  0.400252  & -0.0857 &0.509121  &   -0.1678 & 0.095534 \\
 Pressure&     0.0306 &  0.833578&-0.0013& 0.207932& -0.07 & 0.274788 &-0.1448  &0.116387 \\
 Humidity& -0.1409 &  0.06016 &0.1835&0.002848 & 0.1744 &0.007503  &0.1349 &0.102513 \\
 Light   &0.3074 &  0.00024& 0.0498 & 0.778315& -0.218 &0.010311 &\textbf{-0.3212} &0.0000258 \\
 Audio&  0.1957  & 0.041155 & -0.0196 &0.902739& -0.2103 & 0.022094 &-0.0964 &0.268128\\
\hline \hline
\multicolumn{9}{|c|}{\textbf{Spearman Correlation}}     \\
  \hline
 Temperature  & 0.3093	& 0.002649 	& -0.0749	&0.370261 	& -0.0664 	&0.445534	& -0.1668		&0.075632    \\
 Pressure       &  0.0212	& 0.3078132  	& -0.0366	&0.294657  	& -0.1097 	&0.256761	& -0.1572		& 0.245456    \\
 Humidity        & -0.1877	& 0.01258		& 0.1672	&0.003251	& 0.2646 	&0.003597	& 0.1634 		&0.156731 \\
 Light              & 0.3578	& 0.76863 	& 0.039 	&0.613514	& -0.2542 	&0.021563	& \textbf{-0.4056} &0.0000145\\
 Audio            &  0.2036	&  0.811459 	& -0.0124	& 0.782695 	& -0.2276	&0.012345 	& -0.1112		&  0.116357\\
 \hline
 \hline
\multicolumn{9}{|c|}{\textbf{Kendall Correlation}}     \\
\hline 
 Temperature  	& 0.2572	& 0.001526	& -0.0573	& 0.382667	& -0.0521	&0.491273 	& -0.1296 				& 0.094615  \\
 Pressure		&  0.0192	&  0.836026	& -0.0325	& 0.210126	& -0.0945	&0.265795  	& -0.135 				&  0.118883   \\
 Humidity		& -0.1524	& 0.053868 	& 0.1204	& 0.002813	& 0.1899	& 0.00981	& 0.1203				&0.111096 \\
 Light   		& 0.2793	& 0.000338	& 0.0274	& 0.744178	& -0.1867	&0.010056 	& \textbf{-0.3067}		& 0.0000375\\
 Audio		&  0.1623	& 0.046456 	& -0.0088	& 0.911926	& -0.1748	& 0.021161	& -0.0866				& 0.26605\\
\hline
\end{tabular}%

\end{table*}

\begin{table*}[t!] 
\centering
\caption{Correlation between variables and surveys for Experiment 2 - Group~B.}
\label{ex2tempB}
\begin{tabular}{  |c|c|c|c|c|c|c|c|c|}
 \hline
  & \multicolumn{2}{|c|}{\# of People} & \multicolumn{2}{|c|}{PSQI} & \multicolumn{2}{|c|}{PSS} & \multicolumn{2}{|c|}{K10}   \\
 \hline
& $r$ & $p$-value & $r$ & $p$-value & $r$ & $p$-value & $r$ & $p$-value  \\ \hline
  \hline
  \multicolumn{9}{|c|}{\textbf{Pearson Correlation}}    \\
  \hline
 Temperature  	& -0.1713 		&0.167645		& -0.0213	&0.732692 	& 0.2564			&0.039818	& 0.2256	& 0.092844     \\
 Pressure		&  -0.16 	&0.16604 			& 0.1326	& 0.34003		& 0.1446			&0.184859 	 & 0.1832	& 0.08292\\
 Humidity		& 0.3		&0.02037			& 0.1304	&0.739989 	& -0.1486			& 0.1848 	 & -0.1478	& 0.245409\\
 Light   		& 0.1998 		&0.036166		& 0.0872	&0.373814 	& -0.2917			&0.068944 	 & -0.2814	&0.041359 \\
 Audio		&  0.3242 		& 0.002236		& 0.0305	&0.720883	& \textbf{-0.4785}	&0.0000032	 & -0.25	& 0.006185\\
 \hline \hline 
\multicolumn{9}{|c|}{\textbf{Spearman Correlation}}    \\
  \hline
 Temperature  	& -0.1624		&0.155341 & 0.0655		&0.706657  & 0.2259		&0.043672 & 0.177		&  0.116069 \\
 Pressure		&  -0.1413		&0.16359   & 0.1618		&0.361661 & 0.1509		& 0.164022 & 0.1807		&  0.086878     \\
 Humidity		& 0.284		&0.015714 & 0.0896		&0.773536 & -0.1695		&0.149009 & -0.1486		&  0.241944\\
 Light   		& 0.2542		&0.034651 & 0.0583		& 0.434999 & -0.2066		&0.073265 & -0.241		& 0.044556\\
 Audio		&  0.342		&0.002385  & -0.0604		&0.822361 & \textbf{-0.5251}		& 0.0000006& -0.2953		& 0.009548\\
 \hline  
  \hline
\multicolumn{9}{|c|}{\textbf{Kendall Correlation}}     \\
\hline
 Temperature  	& -0.1269		&0.223645  	& 0.0521		&0.833393 & 0.1711		&0.035864 & 0.1469  &  0.096612 \\
 Pressure		&  -0.1252		& 0.45621  	& 0.1345		&0.32236 & 0.1186		& 0.154325& 0.1582  &0.07123   \\
 Humidity		& 0.2187		& 0.02132	& 0.0638		&0.667623 & -0.1122		&0.155612  & -0.1123 &0.323356\\
 Light   		& 0.1988		& 0.045622	& 0.044		& 0.465674 & -0.1505		&0.054663 & 0.1844 &0.044653\\
 Audio		&  0.2889		& 0.003542 	& -0.0451		&0.676346  & \textbf{-0.3991}	&0.0000087 & -0.2501 &0.006776\\
 \hline
  
\end{tabular}%

\end{table*}

\begin{table*}[t!] 
\centering
\caption{Correlation between variables and surveys for Experiment 3.}
\label{ex3temp}
\begin{tabular}{  |c|c|c|c|c|c|c|c|c|}
 \hline
  & \multicolumn{2}{|c|}{\# of People} & \multicolumn{2}{|c|}{PSQI} & \multicolumn{2}{|c|}{PSS} & \multicolumn{2}{|c|}{K10}   \\
 \hline
& $r$ & $p$-value & $r$ & $p$-value & $r$ & $p$-value & $r$ & $p$-value  \\ \hline
  \hline
  \multicolumn{9}{|c|}{\textbf{Pearson Correlation}}    \\
  \hline
 Temperature  	& -0.0816 		&0.0815			&-0.0831 	& 0.0600		&-0.0466 			&0.8073		& -0.0637		&0.3448      \\
 Pressure		& -0.0201 		& 0.8441			& 0.0809		& 0.3617		&0.0892 			&0.1687 	 	& 0.0536		&0.5484 \\
 Humidity		& -0.0886		&0.0547			& -0.018		& 0.3667		&-0.014 			&0.6137 		 &-0.0215		& 0.0656\\
 Light   		&  -0.0943	       &0.056			&-0.0576 		& 0.589		& 0.0675			&0.3995	 	&0.057		& 0.5624\\
 Audio		&  -0.0744		& 0.043			& \textbf{0.1938}		&0.0017		&\textbf{0.1408}			&0.0007	 	&\textbf{0.1532}	& 0.004\\ 
 \hline \hline 
\multicolumn{9}{|c|}{\textbf{Spearman Correlation}}    \\
  \hline
 Temperature  	&  -0.1029		&	0.0818		&-0.1133 	& 0.0552		& -0.0172			&0.7714		& 0.0568		&0.3375      \\
 Pressure		&-0.0129  		& 	0.8281		& 0.0541		& 0.3607		& 0.0819			& 0.1663	 	&0.035 		&0.5552 \\
 Humidity		& -0.1141		&	0.0535		& -0.0655		& 0.2685		& -0.028			&0.6363 		 &-0.1188		&0.0443 \\
 Light   		&   -0.0932		&0.7251			& -0.0294		& 0.6198		& 0.0414			&0.4852	 	& 0.0352		&0.5531 \\
 Audio		&   -0.1043		& 0.6546			& \textbf{0.1829}		&0.0019		& \textbf{0.194}			&0.001	 	& \textbf{0.1682}		&0.0043 \\
 \hline  
  \hline
\multicolumn{9}{|c|}{\textbf{Kendall Correlation}}     \\
\hline
 Temperature  	&  -0.087		&	0.0832		&-0.0851 	& 0.0645		&-0.0108 			&	0.7083	& -0.0443		&    0.3348   \\
 Pressure		&  -0.0097		& 	0.8913		&0.0451 		& 0.3457		& 	0.0667		& 0.1567	 	& 0.0307		& 0.6484\\
 Humidity		& -0.0994		&	0.7516		& -0.0393		& 0.3761		& -0.0215			& 0.6453		 &-0.0829		& 0.0446\\
 Light   		&  -0.0771			&0.8422			& -0.0232		& 0.5756		& 0.0355			&0.3345	 	& 0.0258		&  0.5443\\
 Audio		&  -0.0983			& 0.8251			&\textbf{0.1435} 		&0.0016		&\textbf{0.152}			&0.0005	 	& \textbf{0.136}		&0.003 \\
 \hline
 
\end{tabular}%

\end{table*}

For Experiment 2,  the strongest correlation available in Group~A is that between the K10 results and light, resulting in a Pearson correlation of $-0.3212$,  a Spearman correlation of $-0.4056$, and a Kendall correlation of $-0.3067$, shown in Table~\ref{ex2tempA}.  In Group~B, this correlation was lower,  for Pearson, Spearman, and  for Kendall, as shown in Table~\ref{ex2tempB}.  However, there is a strong correlation in Group~B for audio.  The correlation between ambient audio and PSS results was a $-0.4785$ Pearson coefficient and a $-0.5251$ Spearman coefficient.  

It is hypothesized at this point that the discrepancy between Group~A and Group~B results can be explained through the data distribution. Figure~\ref{fig:g1light} and Figure~\ref{fig:g2light} indicate that Group~A had greater light values.  Figure~\ref{fig:g1audio} and Figure~\ref{fig:g2audio} indicate that Group~B had a greater audio values. 

Again,  the $p$-values deemed statistically significant, thus disproving the null hypothesis.  The null hypothesis was disproven for several of our variables in relation to surveys, but not always matching for the two groups.

The only two null hypotheses that were disproven for both groups were those discussed as ``moderate" in strength: light versus K10 and noise versus PSS.  So while the strength of the correlation for these two categories does not match in each group, they agree that it can not be said that there is no correlation.


 Table~\ref{ex3temp} shows the results for Experiment 3.   None of the $r$ values meet the threshold for moderate correlation.  In fact, the magnitudes in this experiment are small.  However, we will analyze the strongest of the weak correlations and check which are statistically significant.  
    
    The environmental variable with the strongest $r$ values in this experiment is audio.  Audio shows Spearman correlations of $0.1829$, $0.194$, and $0.1682$ with the PSQI, PSS, and K10 surveys respectively.  This trend is confirmed by the Pearson and Kendall values.  All three of these relationships are statistically significant as confirmed by the Spearman, Pearson, and Kendall $p$-values in Table~\ref{ex3temp}.  This means that while the relationship is weak, it cannot be said that there is no relationship between the two variables.  In summary, this is true for the relationships between stress and noise, and distress and noise/humidity.  It is appropriate to use Spearman correlation analysis here, as normal distribution was observed for the humidity and noise variables.

\section{Conclusion} \label{con}
This paper introduces a framework and presents the results and correlation analysis of an experiment on a system to assess an individual's wellness.  The system uses BLE technology to communicate sensor data from the sensor board to a smartphone application developed for this system.  The application asks the individual to respond to questions from three psychological surveys three times per day over a period of five days.  During the completion of the surveys, BLE communications data for five environmental variables: temperature, humidity, light, noise, and air pressure.  Survey questions were taken from three existing surveys: The PSS measuring stress, the K10 measuring distress, and the PSQI measuring sleep quality.

The experiment presented in this paper had 62 participants collect 530 valid survey submissions with related environmental data.  A correlation analysis was done measuring the Kendall, Spearman, and Pearson correlation coefficients for each environmental variable compared to each of the three psychological surveys.  Overall, the introduced framework was able to capture useful wellness information and according to the experimental results  identify a relationship between light and audio with the stress and distress. However, further experimentation with a larger number of participants and with a greater variety of parameters is necessary. At the same time, the framework might need sensor calibration before starting the experiments.

\section{Ethical Standards}
This study was performed by approval of the appropriate
ethics committee and in accordance with the ethical standards with REB \# 18-04-013 and its later amendments.

\bibliographystyle{IEEEtran}
\bibliography{KateSystemsBib}

\begin{IEEEbiography}
[{\includegraphics[width=1in,height=1.25in,clip,keepaspectratio]{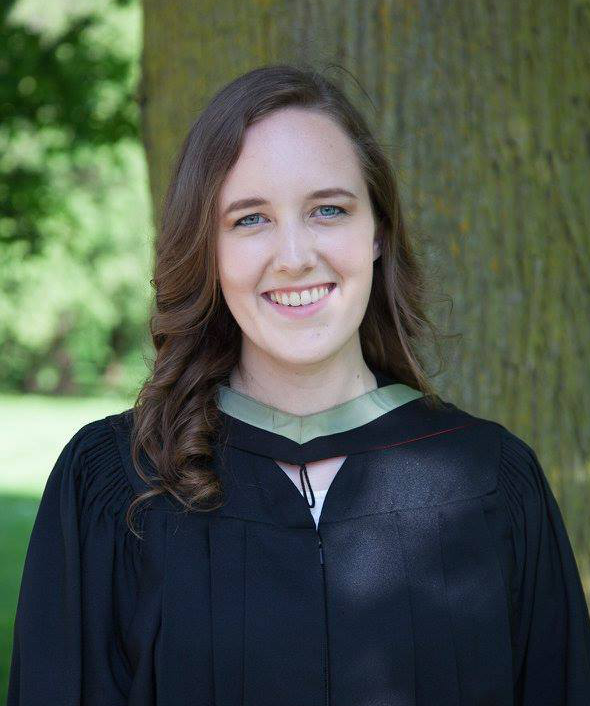}}]{Katherine McLeod} (S'17) received the B.E. degree and the M.A.Sc. degree
both in computer engineering from the School of Engineering at the University of Guelph, Ontario, Canada. Her research interests lie in the area of Internet of Things for healthcare applications with a focus on wellness assessment and smartphone-based applications.
\end{IEEEbiography}

\begin{IEEEbiography}
[{\includegraphics[width=1in,height=1.25in,clip,keepaspectratio]{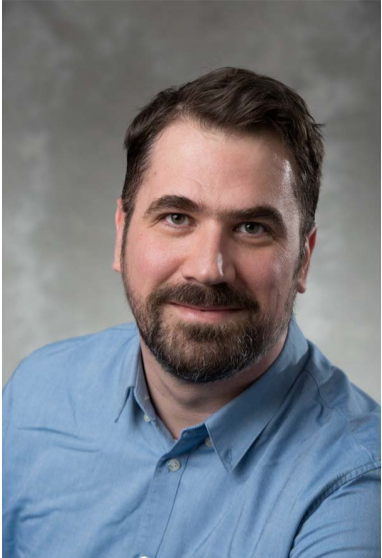}}]{Petros Spachos} (M'14--SM'18) received the Diploma  degree in Electronic and Computer Engineering from the Technical University of Crete, Greece, in 2008, and the M.A.Sc. degree in 2010 and the Ph.D. degree in 2014, both in Electrical and Computer Engineering from the University of Toronto, Canada.  He was a post-doctoral researcher at University of Toronto from September 2014 to July 2015. He is currently an Assistant Professor in the School of Engineering, University of Guelph, Canada. His research interests include experimental wireless networking and mobile computing with a focus on wireless sensor networks, smart cities, and the Internet of Things. He is a Senior Member of the IEEE.
\end{IEEEbiography}

\begin{IEEEbiography}[{\includegraphics[width=1in,height=1.25in,clip,keepaspectratio]{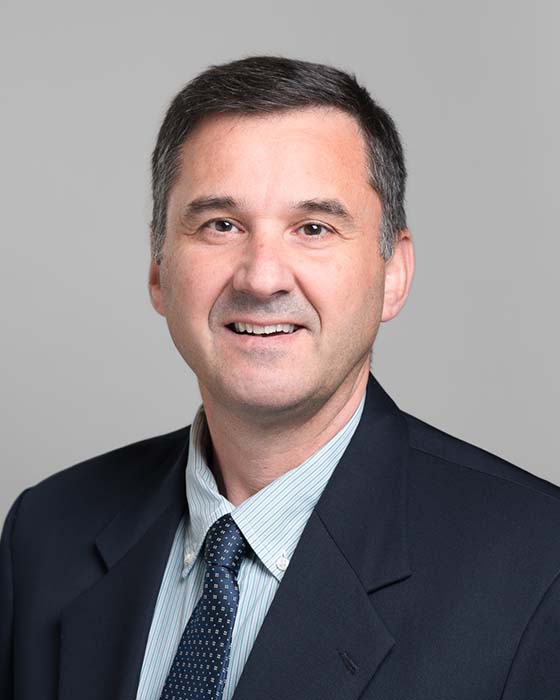}}]{Konstantinos N. (Kostas) Plataniotis}  (S'90--M'92--SM'03--F'12) received his B. Eng. degree in Computer Engineering from University of Patras, Greece and his M.S. and Ph.D. degrees in Electrical Engineering from Florida Institute of Technology Melbourne, Florida. Dr. Plataniotis is currently a Professor with The Edward S. Rogers Sr. Department of Electrical and Computer Engineering at the University of Toronto in Toronto, Ontario, Canada, where he directs the Multimedia Laboratory. He holds the Bell Canada Endowed Chair in Multimedia since 2014. His research interests are primarily in the areas of image/signal processing, machine learning and adaptive learning systems, visual data analysis, multimedia and knowledge media, and affective computing. Dr. Plataniotis is a Fellow of IEEE, Fellow of the Engineering Institute of Canada, and registered professional engineer in Ontario.

Dr. Plataniotis has served as the Editor-in-Chief of the IEEE Signal Processing Letters. He was the Technical Co-Chair of the IEEE 2013 International Conference in Acoustics, Speech and Signal Processing, and he served as the inaugural IEEE Signal Processing Society Vice President for Membership (2014 -2016) and General Co-Chair for the 2017 IEEE GLOBALSIP. He serves as the 2018 IEEE International Conference on Image Processing (ICIP 2018) and the 2021 IEEE International Conference on Acoustics, Speech and Signal Processing (ICASSP 2021) General Co-Chair.
\end{IEEEbiography}

\end{document}